# THE CANADA-UK DEEP SUB-MILLIMETER SURVEY II:
# FIRST IDENTIFICATIONS, REDSHIFTS AND IMPLICATIONS FOR GALAXY EVOLUTION[†]

Simon J. Lilly[1], Stephen A. Eales[2], Walter K.P. Gear[3], François Hammer[4], Olivier Le Fèvre[5], David Crampton[6], J. Richard Bond[7], Loretta Dunne[2]

[1]Department of Astronomy, University of Toronto, 60 St. George Street, Toronto, Ontario M5S 3H8, Canada

[2]Department of Physics and Astronomy, Cardiff University, P.O. Box 913, Cardiff CF2 3YB, United Kingdom

[3]Mullard Space Science Laboratory, University College London, Holmbury St. Mary, Dorking, Surrey RH5 6NT, United Kingdom.

[4]DAEC, Observatoire de Paris-Meudon, 92195 Meudon, France

[5]Laboratoire d'Astronomie Spatiale, Traverse du Siphon, B.P. 8, Marseilles Cedex 12, France

[6]Herzberg Institute of Astrophysics, National Research Council Canada, 5071 West Saanich Road, Victoria, British Columbia V8X 4M6, Canada.

[7]Canadian Institute for Theoretical Astrophysics, University of Toronto, 60 St. George Street, Toronto, Ontario M5S 3H8, Canada

**ABSTRACT**

Identifications are sought for 12 sub-mm sources detected by Eales et al (1998). Six are securely identified, two have probable identifications and four remain unidentified with $I_{AB} > 25$. Spectroscopic and estimated photometric redshifts indicate that four of the sources have $z < 1$, and four have $1 < z < 3$, with the remaining four empty field sources probably lying at $z > 3$. The spectral energy distributions of the identifications, as defined by measurements or upper limits to the flux densities at 8000 Å, at 15 µm, 450 µm, 850 µm and at 6 cm, are consistent with the spectral energy distributions of high extinction starbursts such as Arp 220. The far-IR luminosities of the sources at $z > 0.5$ are of order $3 \times 10^{12}\ h_{50}^{-2}\ L_O$, i.e. slightly larger than that of Arp 220. As with local ultra luminous infrared galaxies, the optical luminosities of the identified galaxies are comparable to present day L* and the optical morphologies of many of the galaxies show evidence for mergers or highly disruptive interactions. Based on this small sample, the cumulative bolometric luminosity function shows strong evolution to $z \sim 1$, but weaker or possibly even negative evolution beyond. The redshift dependence of the far-IR luminosity density does not appear, at this early stage, to be inconsistent with that seen in the ultraviolet luminosity density. Although the computation of bolometric luminosities is quite uncertain, the population of very luminous galaxies that is detected in the surveys at $z > 1$ is already matching, in the far-IR, the bolometric output in the ultraviolet of the whole optically-selected population. Assuming that the energy source in the far-IR is massive stars, this suggests that the total luminous output from star-formation in the Universe will be dominated by the far-IR emission once the lower luminosity sources, below the current far-IR detection threshold, are included. Furthermore, the detected systems have individual star-formation rates (exceeding 300 $h_{50}^{-2}$ $M_O$ yr$^{-1}$) that are much higher than seen in the ultraviolet selected samples and that are sufficient to form substantial stellar populations on dynamical timescales of $10^8$ yr. The association with merger-like morphologies and the obvious presence of dust makes it attractive to identify these systems as forming the metal-rich spheroid population, in which case we would infer that much of this activity has occurred relatively recently, at $z \sim 2$.

*Key words:*    *galaxies: distances and redshifts – evolution – formation – interactions – starburst  infrared: galaxies*

___

# 1. Introduction

The last few years have seen great progress in obtaining a first outline of the evolution of galaxies over a remarkably wide range of cosmic time, corresponding to redshifts $0 < z < 5$. This progress is epitomized by the plot of ultraviolet luminosity density of the Universe, $\mathcal{L}_{uv}(z)$ (Lilly et al 1996, Madau et al 1996, 1998, Connolly et al 1997) which suggests that $\mathcal{L}_{uv}(z)$ rises quite rapidly with redshift to a peak in the $1 < z < 2.5$, declining in a not yet well determined way to higher redshifts. At the higher redshifts, there are still significant uncertainties in the form of $\mathcal{L}_{uv}(z)$ associated with the photometry and the application of the "uv-dropout" selection criteria (see e.g. Sawicki et al 1997). In translating the ultraviolet luminosity density $\mathcal{L}_{uv}(z)$ into a star-formation rate (e.g. Madau et al 1996), there are additional uncertainties regarding the effects of dust obscuration and the initial mass function.

Despite this definite progress, the relationship between the stars produced at different epochs and the present-day morphological components of the galaxy population remains unclear. The evolution to $z \sim 0.8$ appears to be primarily due to relatively small galaxies with irregular morphology and to the disk components of galaxies (see e.g. the morphological analyses of Brinchmann et al 1998 and Lilly et al 1998a, also Guzman et al 1997 and references therein). The nature of the galaxies seen at $z > 3$ (Steidel et al 1996) is still quite controversial (see e.g. Dickinson 1988, Trager et al 1997 and references therein), and very little is really known about the nature of galaxies in the crucial intermediate redshift range $1.5 < z < 3$. A major unsolved problem concerns the origin of the stars seen in the metal-rich spheroid components of galaxies, which, today, constitute a half to two-thirds of all stars (see Fukugita et al 1998). There are inconclusive observational arguments both in favor of rapid and homogeneous spheroid formation at very high redshifts (e.g. Ellis et al 1997) and in favor of a formation through mergers spread over a broad range of epochs (e.g. Kauffmann et al 1996). These alternatives are often viewed as mutually incompatible, although the formation of the bulk of metal-rich spheroid stars in highly dissipational mergers of gas-rich systems at high redshifts would combine the attractive features of both scenarios. A problem, however, with this scenario has been the lack of detection of a substantial population of luminous star-forming high redshift galaxies with the high star formation rates (several × $10^2$–$10^3$ $M_\odot yr^{-1}$) that would be required to produce major spheroidal components of galaxies on typical dynamical timescales ($10^8$ yr).

Until very recently, our view of the evolution of typical galaxies at high redshifts has come almost exclusively from studies of starlight in the optical and near-infrared wavebands. Very little has been known about that component of stellar energy that is absorbed by dust and emerges in the far-infrared waveband in the form of thermally re-radiated emission from dust. Studies of optically selected galaxies at high redshift (see e.g. Sawicki et al 1998, Meurer et al 1998, Pettini et al 1998, Dickinson 1998) suggest that dust obscuration is important.

Studies of the local Universe indicate that between 30-40% of all stellar luminosity emerges as thermally re-radiated dust emission (see e.g. Dwek et al 1998 and references therein). Furthermore, the recently detected far-IR/sub-mm background seen in the COBE data (Puget et al 1996, Hauser et al 1998, Fixsen et al 1998, Schlegel et al 1998) has a $\nu I_\nu$ energy content that is as large or larger than the optical/near-IR background that is obtained by integrating the galaxy number counts (see e.g. Pozzetti et al 1998, Dwek et al 1998). This indicates that dust continues to play a major role at high redshifts and that a half or more of the energy from stellar nucleosynthesis at cosmological redshifts emerges in the FIR (it should be noted that any differences in the redshifts at which these backgrounds are produced will change their relative energy content only as $(1+z)$). In terms of the light from young stars, the balance may be tipped even further in favor of the far-IR because a significant fraction of the optical background will be coming from old stars (note that the energy of the background is three times higher at $K$ than at $U$, see Pozzetti et al 1998).

Understanding the nature and redshifts of the sources responsible for the far-IR/sub-millimetre background is therefore vital for an understanding of galaxy evolution. Key questions are (a) whether the redshift dependence of this obscured star-formation component follows that of the unobscured component studied hitherto, and (b) whether the individual objects emitting in both wavebands are similar, i.e. whether the effects of dust obscuration are uniform across the star-forming population at high redshift.

Several groups (e.g. Smail et al 1997, 1998, Hughes et al 1998, Barger et al 1998 and ourselves) are pursuing deep surveys in the sub-millimetre waveband at 450 and 850 μm with the new SCUBA bolometer array (Holland et al 1998, Gear et al in preparation) on the 15m James Clerk Maxwell Telescope (JCMT) located on Mauna Kea. The spectral energy distributions of galaxies in the sub-millimetre waveband rise sharply with frequency up to a peak in the 70-120 μm range, leading to extremely beneficial *k*-corrections at 850 µm., a situation quite unlike that seen at all wavelengths λ < 100 µm. In consequence, a typical star-burst galaxy (with an effective dust temperature of around 30K and effective emissivity proportional to $\nu^{1.5}$) has a roughly constant observed flux density at 850 µm over the entire $0.5 < z < 6$ redshift range, especially if Ω = 1 (see Fig 4 below). Observations at 850 µm are thus very sensitive to star-formation at very high redshifts.

It has recently become clear (Smail et al 1997, Hughes et al 1998, Barger et al 1998, Eales et al 1998, hereafter Paper I) that a significant fraction (20-30%) of the far-IR background at 850 µm can be resolved into relatively bright sources with $S_{850} \geq 3$ mJy. The surface number density of such sources is almost two orders of magnitude higher than expected in simple "no-evolution" models for the sub-mm sky (Smail et al 1997, Paper I). As pointed out by several authors, these sources must have very high far-IR luminosities, around $3 \times 10^{12}$ $h_{50}^{-2}$ $L_0$, as long as their redshifts lie in the $0.5 < z < 6$ range where the flux density-redshift relation is flat. This is 50% higher than even the very luminous local source Arp 220. Many of these sources undoubtedly lie at high redshifts (Hughes et al 1998, Barger et al 1998, Ivison et al 1998) although the actual redshift distribution is still far from clear on account of the very small samples studied to date, ambiguities in the identifications (see Richards 1998) and difficulties in measuring redshifts (e.g. Barger et al 1998). The sources at high redshift are likely responsible for a significant fraction of the integrated extragalactic background, and therefore must presumably represent a corresponding fraction of all the stellar luminosity that has ever been emitted in the history of the Universe (assuming that stellar nucleosynthetic energy dominates over that released by accretion onto massive black holes). The high individual luminosities (and high implied star-formation rates), coupled with the high total output of stars from the population, makes it attractive to identify these sources as producing many of the stars seen in the present-day metal-rich spheroid components of galaxies.

A difficult question concerns the fraction of this energy that comes from hidden active galactic nuclei. In the local Universe, the evidence (see e.g. Genzel et al 1998) is that AGN provide a significant but not dominant contribution. The observational situation on the high redshift ULIRGS is open (see Ivison et al 1998, Frayer et al 1998). Ascribing a substantial fraction of the energy output of this population to AGN would require a major upwards revision in the total energy output of AGN (but see Haehnelt et al 1998).

In this paper we report and analyse the identifications of the first 12 sub-mm sources (Paper 1) that have detected in the "Canada-UK Deep Sub-mm Survey". This survey at present covers 24.5 arcmin$^2$ of the sky to a 3σ depth of about 3 mJy at 850 µm. It is intended to eventually cover at least 120 arcmin$^2$ to this depth. It should be noted that this survey is therefore wider than, but only half as deep as, the HDF study reported by Hughes et al (1998).

We take $H_0 = 50$ $h_{50}$ kms$^{-1}$Mpc$^{-1}$ and assume $\Omega_0 = 1$ unless otherwise noted. We find that almost all the galaxies have spectral energy distributions that are similar to that of Arp 220. Accordingly, we define the bolometric luminosity $L_{bol}$ as the luminosity integrated over this spectral energy distribution from ultraviolet to radio wavelengths, normalized to the measured brightness at 850 µm. For comparisons with the extensive literature on IRAS sources, it should be noted that this is 1.5 times larger than the $\nu L_\nu$ luminosity of Arp 220 evaluated at 60 µm. It should be noted in passing that the ratio between far-IR (3 µm < λ < 1 mm) and visible (3000 Å < λ < 3 µm) luminosities in Arp 220 is $L_{FIR}/L_{opt} = 33$, so the contribution from the optical is very small. Nevertheless, even in the far-IR the bolometric luminosity of sources detected in the sub-mm is still uncertain because of the poorly defined temperatures and effective emissivities. Correspondingly there could be systematic uncertainties in $L_{bol}$ of a factor of 2-3 at $z \sim 1$.

## 2. Identifications of sub-mm sources

The twelve sub-mm sources in Paper I are the first to be detected in a UK-Canadian sub-mm survey of fields from the Canada-France-Redshift Survey (CFRS, Lilly et al 1995a, Le Fèvre et al 1995 and references therein). The sub-mm survey is carried out with the SCUBA array bolometer (Holland et al 1998) on the UK-Canada-Netherlands 15m JCMT on Mauna Kea. The details of the sub-mm observations and analysis of the number counts are given in Paper 1. As noted there, Monte Carlo simulations and other analyses of the sub-mm data indicate that of order one of the 12 sources may be spurious.

The CFRS fields were chosen not only because of the extensive redshift data available for galaxies with $I_{AB} < 22.5$, spanning the redshift interval $0 < z < 1.3$, but also because these fields have a wealth of deep imaging data from the Canada-France-Hawaii Telescope (CFHT) and the Hubble Space Telescope (HST), deep 7 and 15 µm maps taken with Infrared Space Observatory (ISO) (Flores et al 1998ab) and deep radio images from the Very Large Array (VLA) (Fomalont et al 1992, Hammer et al 1995b). Furthermore, portions of some of the fields have also been studied by others as part of the Hawaii and Steidel et al deep survey programs. While the observations at all wavelengths are not as deep as in the Hubble Deep Field, the fields are much larger allowing us to study a relatively large contiguous area of sky. Our new sub-mm survey is a "field survey" and is not targeted at known objects. It will ultimately cover a good fraction of the 200 arcmin$^2$ of the CFRS–0300+00 (Hammer et al 1995a) and CFRS–1417+52 fields (Lilly et al 1995b).

*2.1  Optical and radio imaging data*

Deep optical images of these fields were searched for identifications of the sub-mm sources. These images included the original $V$, $I$ and $K$ images of the CFRS (Lilly et al 1995a), supplemented by further deep images of these fields that have been obtained as part of the Canada-France Deep Fields program (CFDF, publications in preparation) using data obtained with the UH8k mosaic camera at CFHT (in $V$ and $I$) and the PF camera at the Kitt Peak National Observatory Mayall 4m (in $U$). For many of the sources, HST images are available from our earlier studies of these fields or from the "Groth strip" which bisects the 1417+52 field. These optical images are much less deep than the HDF, although this is not necessarily a problem. The high surface density of faint galaxies that lie below our own limit (i.e. at $I_{AB} > 25$) results in many such galaxies lying within the search radius, making it hard to know which, if any, is the correct identification. Our own study ducks this question by simply classifying these sources as "empty fields".

For the eight sub-mm sources in the 1417+52 field, searches were also made for identifications with radio sources in the Fomalont et al (1992) deep VLA survey and Flores et al (1998a,b) deep ISO images of these fields at 7 μm and 15 μm.

*2.2  Identification probabilities*

Even with a 15m telescope, the beam size (FWHM) at 850 μm is 15 arcsec and at 450 μm (for those sources detected at all) it is 8 arcsec. Thus, until more accurate positions are available from interferometers, identifications of the sub-mm sources at other wavelengths must be based on probabilistic arguments. Our own approach is to do this strictly on the basis of positional coincidence (see e.g. Browne and Cohen 1978, Downes et al 1986), allowing us to subsequently examine the properties of the proposed identifications (such as their multi-wavelength colors) in a way that is independent of the identification procedure.

The identification of the optical galaxies responsible for the sub-mm emission through probabilistic arguments involves a number of subtleties. The probability that the nearest member of a population of objects with surface density $n$ is located within a distance $d$ from a random position on the sky is given by $P = 1-\exp(-\pi n d^2)$. The $P$ statistic has frequently been used for identifying sources (e.g. Downes et al 1986). The quantity $P$ tells us the fraction of sources in a sample of size $N$ that would be expected to have an incorrect candidate identification lying within this distance $d$, i.e. $N_{spurious} (<P) = NP$. This $P$ statistic represents a starting point, but is not what is really required, which is rather *the probability that a particular claimed identification is in fact correct*.

For the radio (and ISO) catalogues, $P$ may be simply computed from the number density $n$ of all sources in the catalogue, irrespective of their radio (or mid-IR) brightness, since $n$ is sufficiently low that any source within the 5 arcsec search radius has a low $P$. On deep optical images, however, the number density of sources is high enough that a 5 arcsec search radius may contain a significant number of random galaxies (at $I_{AB} < 25$ the average density is about 0.8 sources per error circle and at the HDF limit of $I_{AB} \sim 29$, it is about ten times larger). Thus it is advantageous to compute the number density $n'$ of galaxies brighter than the magnitude of the galaxy, i.e. $n' = n(<m)$, decreasing the values of $P$ for the brighter galaxies (since $n$ is much reduced from that of the optical catalogue as a whole). However, it should be noted that this procedure introduces a bias towards underestimating the probability of unrelated sources because many almost independent galaxy samples are being searched simultaneously (i.e. the galaxies brighter than $17^{th}$ magnitude, those brighter than $18^{th}$, those brighter than $19^{th}$, and so on) so it becomes more likely that one of these samples turns up a low-$P$ positional coincidence. The true fraction of randomly located sources with candidate identifications lying within $d$ is $P' = \alpha P$. The correction term $\alpha$ is a function of $P$, the search cut-off radius and the magnitude range over which identifications are sought. $P'$ has been determined using Monte Carlo simulations that matched our operational procedure. It should be noted that the correction term $\alpha$ can be quite large. In our simulations, it is 6.5 for all $P \leq 0.01$, decreasing to higher $P$ ($\alpha \sim 3.5$ at $P \sim 0.1$). It should be noted that in almost all cases, there is either only one object visible within the search radius of 5" or one object has a $P'$ value that is so much smaller than the others that it is much more likely to be the identification.

It should be noted that we have computed the separations $d$ as the distance between the centers of the sub-mm and optical/radio sources and have not considered the size of the optical galaxy. This makes some astrophysical sense in that the energy sources for the sub-mm sources, whether star-burst or AGN, are likely to be located at the centers of the galaxies. This procedure also has the merit of being well-defined and of avoiding ambiguities in defining the size of the optical galaxies. It is possible in principle, but extremely unlikely in practice, that we have missed off-centered sources in nearby large galaxies.

However, as noted above, the real statistical question that we wish to answer is "what is the probability that a particular identification of some individual source, with some value of P (or P'), is actually the *correct* identification?". As noted above, a sample of *N* sources should contain approximately *NP* spurious identifications. Thus, a low value of *P* for any individual source is not, on its own, enough to make an identification secure. Rather, and perhaps counter-intuitively, one has to look at the sample as a whole and determine the number of identifications in the sample (with that value of P) relative to the number of spurious identifications (with that same P) that would have been expected if the sub-mm and optical populations were completely unrelated. Only if this ratio is high, i.e. if there are many sources with this value of *P* compared to the number expected by chance, can a particular individual source be regarded as securely identified.

This is illustrated in Fig 1, where we show the distribution of (optically determined) P values for the identifications in the present program compared with the distribution of P values found in the Monte Carlo simulations. This shows that the three identifications with $P < 0.001$ are very secure. The three identifications with $P \leq 0.05$ are quite secure (especially as one has a $P < 10^{-3}$ from the radio) – although even at this level, one of the identifications could be spurious. The two remaining identifications, with $P \sim 0.01$, while possibly correct, could in fact be chance associations. It should be noted that, apart from the radio sources, it is the brighter optical galaxies at lower redshift (see below) that are the most securely identified. The remaining four sub-mm sources have either no candidate identification visible within the error circle or only a faint candidate with high P', i.e. $P' \geq 0.5$. These are regarded as "empty fields".

The optical and radio positions for all eight identifications are shown in Table 1, together with the associated values of *P* and *P'* (sub-mm positions are given in Paper I). At 850 µm, the identifications are generally located within $1 < d < 3$ arcsec, even for the most securely identified sources (e.g. the radio sources). We regard this as acceptable given the size of the 850 µm beam (FWHM = 15 arcsec) and the low S/N of the detections, the use of blind pointing on the JCMT, and possible mismatches between the astrometric grids, and it is similar to that found by Smail et al (1998). We do not see evidence for any systematic offset between the optical/radio and JCMT astrometric systems (c.f. Richards 1998). There is no trend of decreasing positional offset with increasing faintness of the optical candidate – a potential indicator of spurious identifications (Browne and Cohen 1978). Furthermore, it should be noted how in the four cases where the source was detected at 450 µm the positional discrepancy is reassuringly smaller at the shorter wavelength. The Fig 2 shows a montage of HST images of 7 of the 8 identifications (the other one was not observed with HST).

Positional coincidence offers a simple, robust and conservative approach to making identifications. Problems could obviously arise if the sub-mm sources were not in fact randomly located relative to physically unrelated galaxies – an obvious possibility being if they are gravitationally lensed by foreground galaxies (c.f. the discussion in Hughes et al 1998). Our procedure would tend to identify the bright foreground lens rather than the real background source. Clustering effects could also bias the results although at least in this case the redshifts would be correctly estimated.

It should be noted that our estimates of *P* and *P'* do not take into account other unusual characteristics (i.e. photometric, spectral or morphological) that the correct identifications are likely to exhibit, which could in principle further reduce the effective density *n* and thus reduce *P* and *P'*.

Many of the galaxies claimed as identifications had already been catalogued in the CFRS program and two already had spectroscopic redshift measurements. One bright galaxy was subsequently observed spectroscopically in June 1998 with the OSIS spectrograph on CFHT, yielding $z = 0.074$. For the remainder, we have estimated likely redshifts on the basis of their optical-infrared *UVIK* colours (see Fig 3). Photometric data for the identifications is given in Table 2. The *VIK* colors span a large baseline in color and offer a relatively clean separation between galaxies at $z < 1.5$ and $z > 1.5$. For the latter sources, detection in *U* limits the redshift to $z < 3$. An additional advantage is that the effect of reddening is to move galaxies in the (*V-I*)/(*I-K*) plane in a direction that is largely orthogonal to the direction produced by varying the redshift (see Fig 3).

The following represent notes on individual sources, in order of decreasing reliability of the identification. Photometric data for the identifications is given in Table 2.

*2.3 Identifications and redshifts*

*2.3.1 The six sources considered most securely identified*

<u>CFRS-10B</u> is securely identified with CFRS-10.1141, a bright $I_{AB} = 17$ spiral galaxy with a newly determined spectroscopic redshift of $z = 0.074$. The spectrum shows strong Hα with EW$_0$ ~ 26 Å. Our CFHT and HST images show a relatively normal looking face-on spiral galaxy. It should be noted that the far-infrared luminosity of this source is a factor of ten smaller than that of the others in the sample, but is still substantial, $L_{bol}$ ~ $2\times10^{11}$ $h_{50}^{-2}$ $L_O$.

<u>CFRS-14F</u> is the faintest sub-mm source in the sample but nevertheless one of the most securely identified. It is identified with CFRS-14.1139, an $I_{AB} = 20.5$ galaxy with measured $z = 0.660$ (Lilly et al 1995b). This galaxy is itself securely identified (Hammer et al 1995a) with a 79 μJy radio FKWW–15V.24 (Fomalont et al 1992) and is also detected at 7 μm and 15 μm by ISO (Flores et al 1998ab). The HST images show a dramatic and chaotic morphology indicative of a major merger event. The spectrum shows moderately strong [OII] 3727 in emission (EW$_0$ = 13 Å) and the Balmer lines (Hδ and Hε) in absorption. The optical-infrared colours are of a relatively unobscured late-type galaxy and the large FIR luminosity doubtless comes from a highly embedded region.

<u>CFRS-14A</u> is the brightest sub-mm source in the sample. It is 2" from the 44 μJy radio source FKWW–V15.18 which is itself identified (Hammer et al 1995a) with a faint galaxy. This galaxy is thus a secure identification despite the faintness of the optical galaxy. As noted by Hammer et al., the galaxy is very red and the source was identified by them only in the $K$-band at $K_{AB}$ ~ 19. The galaxy is detected on our F814W HST images and can also be made out on our ground-based $V$-band and $I$-band images (especially once the new CFDF data is added). The galaxy has $I_{AB}$ ~ 24. The *VIK* colors suggest a redshift $z$ ~ 2. There is a 2σ photometric detection at *U*, requiring $z < 3$. The HST morphology is rather indistinct, being quite compact, but even so, it is not completely symmetrical.

<u>CFRS-10A</u> is identified with CFRS-10.1153, an $I_{AB} = 21.5$ galaxy with previously measured $z = 0.550$ (Le Fèvre et al 1995b). The optical spectrum shows only absorption features (but does not extend as far as Hα), but the *VIK* colors indicate a color excess of 0.5 magnitudes in (I-K) relative to an early type galaxy at this redshift. The HST morphology is unspectacular (it was classified by Brinchmann et al 1998 as a mid-spiral) but the galaxy has a strong bar-like feature with a prominent secondary maximum at one end. There is also some evidence for irregular outer isophotes, so it is possible that this is some kind of merger, perhaps viewed from an unfavorable angle.

<u>CFRS-03A</u> is likely associated with the $I_{AB}$ ~ 23.4 galaxy catalogued as CFRS 03.1055. The positional discrepancy (2.8 and 2.0 arcsec at 850 and 450 μm respectively) is acceptable and only the faintness of the galaxy increases *P'* to 0.14. The F814W HST image shows a flattened system (possibly an edge on disk) with multiple components in the center. There is again evidence for some asymmetry in the outer isophotes. There is at present no *K* photometry, or optical spectroscopy, on this object, but the *UVI* colours suggest that this galaxy has the colors of a mid-spiral at $z$ ~ 1.5. It must be at $z < 3$ on the basis of a good detection at *U*.

<u>CFRS-10D</u> is almost certainly identified with CFRS-10.1167, an $I_{AB} = 22.6$ galaxy. Although *P'* ~ 0.15 (the separation from the 850 μm position is 2.8 arcsec and there is no 450 μm position), the F814W HST morphology is very unusual with two high surface brightness components separated by 0.9" with surrounding asymmetric nebulosity that is suggestive of tidal features. The morphology is highly indicative of some form of merger or strong interaction, and this substantially increases our confidence in the reality of this identification. The redshift is estimated from the *VIK* colors as $z$ ~ 0.8, almost irrespective of the degree of reddening.

*2.3.2 The two less secure identifications*

Two sources have candidate identifications with faint galaxies. The values of *P'* for these galaxies suggest that there is a significant chance that these are not individually associated with the sub-mm sources.

<u>CFRS-14C</u> is plausibly identified (*P'* ~ 0.22) with a faint galaxy with $I_{AB}$ ~ 24 located 2" from the nominal 850 μm position. The *VIK* colours lie in the area of $z = 2 – 3$ galaxies almost irrespective of reddening. The good detection in the *U*-band at $U_{AB}$ ~ 25.8 constrains the redshift of this galaxy to be $z < 3$. There is no HST image available and the morphology on the CFHT images is unremarkable.

<u>CFRS-14B</u> is 2" away from a faint galaxy with $I_{AB}$ ~ 25 (*P'* = 0.32). The HST morphology of this faint galaxy is compact but distinctly asymmetric.. The colours are similar to the possible identification of CFRS-14C, albeit a magnitude fainter, leading to a similar estimated redshift. The *U*-band photometric detection is more marginal (about 1.5σ), so the $z < 3$ constraint is correspondingly slightly weaker and the redshift less certain.

*2.3.4 The four empty fields*

The remaining four sources, CFRS-10C, CFRS-14D, CFRS-14E, CFRS-14G, are empty to the reliable limits of our ground-based images, i.e. $I_{AB} > 25$, and in any case the density of sources at such optical levels is rising sufficiently that the reliability of such identifications would be highly questionable We therefore regard these four sources as "empty fields". Nature conspires to make very high redshift sources least detectable in the visible, ISO and radio bands, and at 450 μm, and so these faintest optical sources would not be expected to be detectable at these other wavelengths, and indeed aren't. With no supporting detections at other wavelengths, it is always possible that one or more of these sources is spurious (see the discussion in Paper 1), but most of them are probably real.

*2.4 Discussion: properties of the identifications and empty fields*

Having made the identifications on the basis of positional coincidence alone and estimated the redshifts of the galaxies from spectroscopic observations or from photometric colors, we can now examine the properties of the spectral energy distributions (SEDs) of the galaxies as defined by the flux density ratios between the visible, mid-infrared (15 μm), sub-mm (450 μm and 850 μm) and radio (5 GHz) wavebands. Because of the large uncertainties in the redshifts of many of the galaxies, the SEDs are studied by means of color-redshift relations (comparing the galaxies to a series of standard SEDs) rather than by attempting to construct rest-frame SEDs. We have computed "colors" – defined as $\log(S_1/S_2)$ – between 850 μm and the observed *I*-band, 15 μm, 450 μm, and 5 GHz wavebands for the six SEDS given by Schmitt et al (1997), using modelled interpolations between the points given in that work. In particular, the sub-mm SEDs at $\lambda > 100$ μm are constructed using a 30K radiation component with an effective emissivity ε $\propto \nu^{1.5}$. We have also constructed a final template SED representing Arp 220 interpolating between the photometric measurements given by Rowan-Robinson and Efstathiou (1993) and references therein. These SEDs exhibit a wide range of the ratio between far-IR and visible luminosities.

*2.4.1 The sub-mm luminosities of the galaxies.*

As expected from the flat flux-density redshift relation, the 850 μm flux densities do not correlate with the estimated or measured redshifts. This is shown in Fig 4, which also shows the expected track of Arp 220 (with no evolution) under two different cosmologies. The lack of correlation suggests that the redshift distribution of our sample is unlikely to be affected by any incompleteness in detecting sources above the nominal flux density limit (see Paper I) since any missing sources are unlikely to be biassed towards any particular redshifts.

Fig 4 makes the point (see also Barger et al 1998) that any sub-mm source that is detected at the $S_{850} > 3$ mJy level at $z > 1$ is likely to have a bolometric luminosity that is significantly larger than that of Arp 220 ($z = 0.018$), widely regarded as the archetypal ultra-luminous infrared galaxy (ULIRG) in the local Universe.

*2.4.2 The optical luminosities of the galaxies.*

In contrast, the $I_{AB}$ magnitudes of the identifications do correlate well with the estimated and measured redshifts (see Fig 5). This suggests that the identified galaxies are generally luminous optical galaxies with roughly L* or greater optical luminosities. As seen in Fig. 2, many of the galaxies have highly disturbed or multiple morphologies. These are both characteristics of ULIRGs at low redshifts (see Sanders and Mirabel 1996 and references therein). The empty field sources are consistent with either being similar galaxies at very high redshifts $z > 3$, or less luminous systems at lower redshifts.

*2.4.3 The 0.8-850 μm colors*

For all sources we have measurements or limits to the *I*-band magnitudes. The (0.8-850) colors, defined as $\log(S_{0.8}/S_{850})$ and shown in Fig 6. These track well the colors expected for Arp 220 (and for the highly obscured star-burst SED from Schmitt et al 1997).

The limits to the empty field sources are consistent with these being similar objects at any $z > 2$, and they could be at even lower redshifts if they were more heavily obscured and consequently had higher $L_{FIR}/L_{opt}$.

*2.4.3 The 450-850 μm colors*

The 450 μm and 850 μm bolometer arrays on SCUBA operate simultaneously and so measurements or limits to the 450 μm flux density are available for all sources except one (CFRS-14G is near the edge of the field and the 450 μm camera has a slightly smaller field of view). The (450-850) color is often regarded as a good redshift indicator and, in contrast to the HDF sample of Hughes et al (1998), 4 of our 12 sources are detected at 450 µm. Gratifyingly, these are *four of the five sources with the lowest measured or estimated redshifts.* The measurements and limits track well the predicted (450-850) "colors", as defined above, for the standard template SEDs (Fig 7), assuming that the estimated redshifts are roughly right. Lowering the effective emissivity to $\epsilon \propto \nu^1$ (see e.g. Dunne et al, in preparation) would make all of the sources consistent. It can be seen that, with this quality of data, the three empty field sources with 450 µm upper limits are consistent with lying at almost any redshift.

### 2.4.4 The 15-850 µm colors

The (15-850) colors were extensively used by Hughes et al (1998) in their analysis of the HDF. The $S_{15}/S_{850}$ ratio drops strongly with redshift since the bands lie on either side of the thermal peak at $\lambda \sim 100$ µm. Unfortunately, deep 15 µm data is only available at the moment for the seven sources in the 1417+52 field, only one of which was detected. The one sub-mm source detected by ISO at 15 µm is at low redshift ($z$ = 0.66) and the limits to the colours of the remaining sub-mm sources are again consistent with the Arp 220 and SBH SEDs at their estimated or measured redshifts (Fig 8). The empty field sources have limits to the colors that are consistent with these SEDs at any $z >$ 1.5.

We can also look at the properties of the 15 µm sources that were detected by ISO but which were *not* detected in the sub-mm survey. There are five such sources, all with measured or estimated redshifts in the $0.7 < z < 1.1$ range (Flores et al 1998b). The lower limits to the $S_{15}/S_{850}$ flux density ratio for these sources are consistent with their having the same SEDs at these lower redshifts (Fig 8). They are probably not detected at 850 µm simply because they have slightly lower luminosities.

### 2.4.5 The 850-radio colors

In the CFRS–1417+52 field, we have measurements or limits to the flux densities at 5 GHz (Fomalont et al 1992, Hammer et al 1995). Fig 9 shows the 850 µm to 5 GHz colors of the seven µJy-level radio sources that were detected by SCUBA. The colors of the detected sources track very well the Arp 220 SED and the lower limits for the remaining 5 are not too discrepant.

Interestingly, the two radio sources detected by SCUBA have the steepest radio spectra (consistent with star-formation) whereas the remaining four radio sources in our surveyed area that were not detected in the sub-mm have flat or inverted radio spectra (more indicative of AGN).

### 2.5 Summary: The nature of the sources and their redshift distribution

From the above discussion the following can be concluded:
1. The eight identified galaxies are optically luminous galaxies (comparable to present-day L*) spanning a broad range of redshifts $0.08 < z < 3$ as measured or estimated from their optical colors.
2. For all the identified sources, the broad SED as defined from the optical through to the radio (from measurements or limits at 0.8 µm, 15 µm, 450 µm, 850 µm and at 5 GHz) is consistent with the measured/estimated redshift and a rest-frame SED that matches that of Arp 220 or the heavily obscured starburst SED from Schmitt et al (1997). This SED has a luminosity ratio of approximately 35 between the far-IR and the optical.
3. With the present limited data, the observed properties of the currently unidentified empty field sources would be broadly consistent with those of the identified galaxies if placed anywhere over a wide range of redshifts, $2 < z < 10$. Redshifts as low as $z \sim 1$ are not excluded by the present data but would require an even higher $L_{FIR}/L_{opt}$.
4. The redshift distribution of the sub-mm sources is thus broad, extending from very low $z$ to redshifts that are likely to reach $z \sim 3$ or greater. However, at least three (and probably four) of the 12 sources have $z < 1$ (and these are amongst the most securely identified sources in the sample). Based on the *U*-band detections of the faintest identified sources, it is quite likely that no more than four of the 12 sources have $z > 3$. The median redshift of the sample as a whole is thus unlikely to be much larger than $z \sim 2.5$. The implied redshift distribution is shown in Fig 10a, and compared with the smaller HDF sample of Hughes et al (1998) . In broad terms they appear consistent (but see also Richards 1998). The larger lensed sample of Smail et al (1998) does not have redshift estimates (except for constraints based on detection in *B* or *V*) but appears to have a similar distribution in $I_{AB}$ magnitude (Fig 10b)..

### 3. **High luminosity sources at high redshifts: the hidden phases of galaxy evolution and the relation to the optical picture**

It has been clear since the initial results of Smail et al 1997 (see also Paper 1) that the sub-mm source counts require evolution in the far-IR population. Clearly a key question is to what extent the implied redshift evolution in the sub-mm population is similar to, or different from, that inferred from optical studies (Lilly et al 1996, Madau et al 1996, 1998). As noted in Paper 1, the sub-mm source number counts can in fact be successfully matched by a local 850 μm luminosity function undergoing luminosity evolution such that the luminosity density has the same redshift dependence as in the optical (Madau et al 1996). Furthermore, this model also reproduces (Paper 1) the spectral shape of the FIRAS background if the effective emissivity of the population is relatively flat, i.e. near $\nu^1$.

### 3.1 The redshift distribution – similarities with the optical picture

In this section we examine the implications of the direct redshift information obtained from the optical identifications reported above. The most obvious discrepancy between recently published IRAS-based models (Blain et al 1998) and the present data is the relatively large fraction of sources found in the present study at $z < 1$. As noted above, there are at least two and probably three sources in our sample at $0.5 < z < 1.0$, as well as a single less luminous galaxy at $z = 0.076$ It should be noted that two of the three $0.5 < z < 1.0$ sources are very secure identifications with supporting evidence (in the shape of highly disturbed or multiple morphologies and/or radio and ISO detections) that they are indeed the associated with the sub-mm sources. None of the models considered by Blain et al (1998) appear to have more than about 1% of the sources at these flux densities at $z < 1$. It should be appreciated though that models for the 850 μm sky are still quite uncertain, being generally based (e.g. Blain and Longair 1993, Eales & Edmunds 1997, Blain et al 1998) on extrapolations from the IRAS 60 μm population. This represents a factor of 13 in wavelength straddling the thermal peak at $\lambda \sim 100$ μm, and this extrapolation is therefore highly sensitive to assumptions about the temperature and effective emissivity of the dust. In the near future, SCUBA observations of nearby optically-selected and IRAS-selected galaxies (e.g. Dunne et al in preparation) will lead to a big improvement in the models.

#### 3.1.1 Constraints on the cumulative bolometric luminosity function

In order to look more directly at the implications of the redshift distribution inferred for our sample, we have constructed the cumulative bolometric luminosity function (CBLF) for sources in our sample. The primary motivation is to compare samples selected at very different wavelengths, and it should be remembered that the bolometric luminosities are uncertain by a factor of at least two. Given the limited redshift information and these considerable uncertainties in estimating the bolometric luminosities, *this exercise is intended to be primarily illustrative*. The calculation is done for $\Omega = 1$ but the offset in other cosmologies largely parallels the local luminosity function, so choice of $\Omega$ is in the end largely immaterial given the other uncertainties. The CBLF for our sources has been computed in three redshift bins: (i) $0.5 < z < 1.0$; (ii) $1.0 < z < 3.0$; and (iii) $3.0 < z < 8.0$. It should be recalled that the flux density redshift relation is almost flat between $0.5 < z < 8.0$ (Fig 4) so flux density limited samples will approximate volume limited samples within these broad redshift bins. These redshift bins are chosen so that bin (i) represents objects for which we have good redshift constraints from our spectroscopy and from photometric redshifts of bright galaxies; bin (ii) represents fainter identified sources but which are constrained to $z < 3$ from their detections in the U-band. The four empty field sources for which we have essentially no redshift information are then alternatively assigned either to the very high redshift bin (iii) or to bin (ii).

As noted above, the bolometric luminosities for these sources are estimated by integrating over the Arp 220 SED nomalized to the observed 850 μm flux density, a procedure that may be uncertain at the factor of 2-3 level. In the 0.5 < z < 1.0 bin there are two sources with $L_{bol} > 2 \times 10^{12}\ h_{50}^{-2}\ L_0$ (the third, CFRS–14H, falls below this limit). These are shown in Fig 11 with a Poisson error bar. In the next bin, there are between 4 and 8 sources (depending on whether the 4 empty field sources are included) with $L_{bol} > 3 \times 10^{12}\ h_{50}^{-2}\ L_0$. This range, with associated Poisson error bars, is shown in Fig 11. Finally, there are between zero and 4 sources at $z > 3$ (depending on the redshifts of the empty field sources), again with $L_{bol} > 3 \times 10^{12}\ h_{50}^{-2}\ L_0$. These are shown as an upper limit in Fig 11. This last bin reaches to $z \sim 8$, since this is the maximum accessible redshift for such sources. If all such sources were at the lower end of this bin, due to evolutionary effects, then the CBLF at $z \sim 3$ would effectively be raised by the ratio of volumes (e.g. a factor of two for $3 < z < 5$ for $\Omega = 1$). These estimates of the CBLF are compared in Fig 11 with that of local IRAS galaxies given by Soifer et al (1987) and Saunders et al (1990), which differ only at high luminosities, correcting for the different $H_0$ and different definitions of $L_{bol}$ used in these analyses. We also show (and discuss in Section 3.2), the CBLF of the ultraviolet selected sample given by Dickinson (1998), computing bolometric luminosities in this case as $\nu L_\nu$ at 1500 Å. It should be noted that the vertical normalization of this luminosity function is based on the ground-based samples and is lower than that seen in the (very small) HDF.

Looking at the far-IR CBLF in Fig 11, there is a need for quite strong evolution to $z \sim 1$. In terms of density this is a large factor, about 300, but the CBLF at low redshift is so steep that, viewed in terms of luminosity, the effect is smaller — possibly as low as a factor of 10. This number is, of course, quite uncertain, as it effectively relies on the validity of the computation of bolometric luminosity based on the Arp 220 SED. These could easily represent a factor of 2-3 uncertainty. However, a luminosity change of a factor of ten to $z \sim 1$ is of the same order as the changes in the ultraviolet luminosity density of the Universe to $z \sim 1$ (Lilly et al 1996), i.e. $(1+z)^{3.5}$.

Depending on the redshifts of the empty field sources (and on the slope of the CBLF over the range $2 \times 10^{12} < h_{50}^{-2} (L_{bol}/L_0) < 3 \times 10^{12}$) the evolution at higher redshifts is quite uncertain. However, there is little evidence at this stage for a further dramatic increase in the CBLF at $z > 1$. Putting all the empty field sources at $1 < z < 3$ would allow an increase of a factor of up to 3 from $0.5 < z < 1.0$, but would then require a drastic reduction at $z > 3$. If the empty field sources are at $z > 3$, as we suspect, then the data are consistent with a roughly flat redshift dependence of the CBLF at $z > 1$. *The main point to stress is that the fact that only a third of the sources can lie at $z > 3$ already constrains the CBLF at those redshifts to be comparable to that at $z \sim 1$. We return to this point below.*

### 3.1.2 Constraints on the bolometric luminosity density as a f(z)

In the same way, it is easy to calculate the FIR bolometric luminosity density, $\mathcal{L}_{FIR}$, above $2 \times 10^{12} h_{50}^{-2}$ L$_0$ for $0.5 < z < 1.0$ and above $3 \times 10^{12} h_{50}^{-2}$ L$_0$ at higher redshifts. These are shown compared with the ultraviolet bolometric luminosity density in Fig 12. Faint blue galaxies typically have a spectral index $f_\nu \propto \nu^{-1}$, so the bolometric luminosity density in the ultraviolet has been computed simply using $\nu \mathcal{L}_\nu$ using various estimates of $\mathcal{L}_\nu(z)$ over 1500 Å $< \lambda <$ 2800 Å, themselves based on spectroscopic and photometric redshifts (Lilly et al 1996, Connolly et al 1997, Treyer et al 1998, Madau et al 1998, Sawicki et al 1997).

In terms of a global star-formation history, *this diagram should be interpreted with extreme caution.* There are several issues. In the first place, individual bolometric luminosities are uncertain at least at the factor of two level. More importantly, we know that only 20% of the 850 μm sub-mm background is represented in the far-IR points in Fig 12, so an estimate of a total luminosity density, integrated over all luminosities, would on average be higher, by up to a factor of 5 while, in contrast, the ultraviolet luminosity density estimates already approximate "total" estimates. Also, because the luminosity threshold is lower in the $0.5 < z < 1.0$ far-IR bin, this point is biased high relative to the two higher redshift far-IR points, probably by a factor of about 2. Finally, whereas the far-ultraviolet bolometric luminosity density (particularly as sampled at $\lambda < 2000$ Å) is dominated by young stars, some fraction of the far-IR bolometric luminosity density (particularly at the lower luminosities "missing" from Fig 12) will be due to energy emitted by older stellar populations and some may be due to nuclear (AGN) activity. We have no real way of addressing the question of AGN contamination at this stage. In the local Universe it is likely (Genzel et al 1998) that stellar emission dominates the energy of ULIRGs (with a significant contribution from AGN). There are mixed indications from the few observations available at high redshift (Ivison et al 1998, Frayer et al 1998)

With all these very important caveats in mind, two important points are worth noting. First, even though we are so far considering those brightest sources which produce only 20% of the background (Paper 1) – we refer to this loosely as the "top 20%" of the background – the bolometric luminosity density in the far-IR due to "obscured" star-formation activity *already* appears (with our assumption of an Arp 220 SED) to be comparable to all that seen "unobscured" in the ultraviolet (see also Hughes et al 1998). When the extra star-formation seen in the far-IR at lower luminosities is added (plausibly raising the luminosity density by a factor of up to 5), it is likely that most of the energy from star-formation at $1 < z < 3$ emerges in the far-IR. Second, there is no evidence, as yet, that the *redshift dependence* of the luminosity density in the far-IR is different from that seen in the ultraviolet. We look at this further in the next section.

### 3.1.3 The production of the integrated background in redshift space.

Another way to look at the redshift distribution of the sources is in terms of the redshift distribution of the light in the extragalactic 850 μm background. For an idealized population of star-forming galaxies whose spectral energy distribution is flat (i.e. $f_\nu \propto \nu^0$) and dominated by young stars, it is easy to show (following Lilly and Cowie 1987) *that the cumulative distribution in redshift of the light in the extragalactic background is formally identical to the cumulative distribution in redshift of the production of stars in the Universe.* This result is independent of the cosmology, the nature of the star-formation (i.e. constant or sporadic), possible number non-conservation of galaxies and so on. However at 850 μm, we have a spectral energy distribution that rises as $\nu^{3.5}$ (i.e. for emissivity $\propto \nu^{1.5}$ in the Rayleigh-Jeans tail) so there is *a strong weighting of high redshift star-formation activity in the production of the 850μm background*. The weighting will be simply $f_\nu(\nu_{em})/f_\nu(\nu_{obs})$, or $(1+z)^{3.5}$ over much of the redshift range of interest $0 < z < 6$.

Our data suggests that two-thirds or more of the light in the top 20-30% of the 850 μm background (Eales et al 1998) is emitted at $z < 3$, implying that at least 15% of the whole background is emitted at $z < 3$.

In Fig 13, where we have computed the redshift distribution of the background light for a number of different star-formation histories, assuming that the energy of this star-formation emerges with the spectral energy distribution of an obscured star-burst, e.g. Arp 220 (see above). The distribution of observed light at $S_{850} > 2.8$ mJy in our identified source sample is also shown. This does not reach unity because the unidentified sources have been omitted – their redshifts are unconstrained, but it is assumed for this purpose that they have $z > 2.5$. The contribution from the two less securely identified galaxies estimated to lie at around $z \sim 2.5$ is shown as a dotted line.

As is clear on Fig 13, the weighting effect of the *k*-corrections is such that a galaxy formation/evolution scenario in which 50% of all dust enshrouded star-formation in the Universe occurred prior to $z = 3$ would predict that 85% of the 850 μm background had been produced at $z > 3$! Such an evolutionary scenario would represent a picture in which all spheroidal stars formed very early in the Universe. Even if we *make no assumption* about the redshifts of the fainter sources with $S_{850} < 3$ mJy, our observations would already appear to require at least 15% of the 850 μm background to be produced at $z < 3$. Even though we are sampling just the top 20-30% of the background (Paper 1), we are close to ruling out this scenario which would only be allowed if all the sources at $S_{850} > 3$ mJy lay at $z > 3$ (a rather unlikely situation).

We now make the *assumption* that the redshift distribution of fainter sources with $S_{850} < 3$ mJy follows that at $S_{850} > 3$ mJy. This is a plausible, but not watertight, assumption given the flatness of the 850 μm flux density-redshift relation and it should be noted that the models of Blain et al (1998) have a redshift distribution that only weakly depends on flux density – see the caption to their Fig 13. With this assumption, which must be speculative at present, our results then suggest that the great bulk of obscured star-formation in the Universe occurred at redshifts $z << 3.0$. While this analysis can not be regarded as conclusive until we penetrate deeper in to the background, Fig 13 suggests that the cumulative production of the 850 μm background appears to follow well the expectations of models in which the star-formation peaks in the $1.2 < z < 2$ range and falls thereafter, as is also indicated by analysis of the ultraviolet luminosity density in the Universe (Lilly et al 1996, Madau et al 1996,1998, Connolly et al 1997, see also Sawicki et al 1997).

### 3.2 *Individual luminosities and space densities: differences with the optical picture.*

The analyses above suggest that the redshift dependence of the star-formation producing the far-IR/submm background may be similar to that inferred in the optical, or at least, *that there is no compelling evidence at this time for any differences.* On the other hand, it is clear from Fig 11 that the individual sub-mm sources are much more luminous than their optically selected counterparts. The characteristic bolometric luminosity (in the ultraviolet) of the optically selected samples at $z = 3$ (computed for $\Omega = 1$ from $\nu I_\nu$ at 1500 Å) is about $6 \times 10^{10}$ $h_{50}^{-2}$ $L_O$ (Dickinson 1998), a factor of 50 below the sources detected in the sub-mm in the present survey. But these sub-mm selected ULIRG galaxies have number densities of order $3 \times 10^{-5}$ $h_{50}^{3}$ Mpc$^{-3}$ at $z \sim 2$, considerably less than the $3$-$5 \times 10^{-4}$ $h_{50}^{-3}$ Mpc$^{-3}$ of the L* ultraviolet selected galaxies. Thus, the sub-mm sources are rarer but very much more luminous than the optically selected sources.

### 3.3 *The formation of the metal-rich spheroids?*

The identification of a population of galaxies at high redshift that are producing a substantial fraction of present day stars in high luminosity systems is important because it is then attractive to identify these as producing the spheroidal components of galaxies (or possibly old Pop I disks). Local ultra-luminous IR galaxies have long been proposed as being triggered by major mergers and resulting in the production of massive spheroids (see Sanders & Mirabel 1997 and references therein). The high individual luminosities and implied star-formation rates are consistent with making substantial stellar populations on dynamical timescales.

The population revealed in the sub-mm surveys at high redshift has a sufficiently high number density (more than 100 times higher than in the present-day Universe – Fig 11) that they must be responsible for producing a significant fraction of all stars that have been formed in the Universe (see Fig 12). Not least, the high luminosity sub-mm galaxies detected in the present survey at $S_{850} > 3$ mJy (i.e. with luminosities $L_{bol} >$ $3\times10^{12}$ $h_{50}^{-2}$ $L_O$) are responsible for producing, in the far-IR, a bolometric luminosity density in the Universe that is as large as the bolometric luminosity produced in the ultraviolet by the whole population (integrated over essentially all luminosities) – Fig 12. So, the population sampled by these sources is presumably responsible for the generation of at least 20% of all stars produced in the Universe – a fraction that would presumably rise rapidly as lower luminosities are considered. It should be recalled (e.g. Fukugita et al 1998) that the spheroids contain a half to two-thirds of all stars in the Universe.

The combination of the high integrated production of stars, the high star-formation rates, the incidence of merger-like morphologies and the obvious presence of substantial amounts of dust, make it attractive, though still speculative, to associate these galaxies with the production of the metal-rich spheroid component of galaxies. In this case, these first data from our survey suggest that much of this activity, conservatively at least 50%, has happened at relatively recent epochs, i.e. $z < 3$ (Figs. 12 and 13). We suspect that this could be equally well be accomodated within either hierarchical or monolithic collapse scenarios of galaxy formation.

## 4. Summary

We have searched for identifications of the 12 sub-mm sources ($S_{850} > 2.8$ mJy) reported in Paper 1 at optical, mid-infrared and radio wavelengths with the following results:

1. Six sources can be securely identified, two have probable identifications and four remain unidentified with $I_{AB} > 25$ (for such faint galaxies the chance of random positional coincidences also becomes quite large).
2. Three of the identifications have measured redshifts ($0.08 < z < 0.66$) and redshifts for the remainder have been estimated from *UVIK* photometry. Based on these, four of the identifications are likely to have $z < 1$, with the remaining four identifications likely lying in the $1.5 < z < 3.0$ redshift interval.
3. The spectral energy distributions as defined by measurements or upper limits to the flux densities at 8000 Å, at 15 μm, 450 μm, 850 μm and 6 cm are consistent with the spectral energy distributions of high extinction starbursts such as Arp 220. The bolometric luminosities of the sources at $z > 0.5$ are of order $3 \times 10^{12}$ $h_{50}^{-2}$ $L_O$, i.e. slightly larger than that of Arp 220. The lowest redshift source at $z \sim 0.08$ is about a factor of ten less luminous. As with local ULIRGs, the optical luminosities of the identified galaxies are comparable to present day L* and the optical morphologies of many of the galaxies show evidence for mergers or highly disruptive interactions. Thus in all respects studied so far, the individual sources appear to be similar to the most extreme star-bursts seen locally.
4. Our unidentified sources could be similar galaxies anywhere in a broad range of redshifts at $z > 2$ (possibly extending to very high redshifts). But they could also be at redshifts as low as $z \sim 1$ if they have significantly higher $L_{FIR}/L_{opt}$ ratios.
5. The bolometric luminosity function at $0.5 < z < 1.0$ shows strong evolution relative to that defined by local IRAS galaxies at 60 μm, although the comparison is rendered imprecise by the large difference in wavelengths. While the strength of this evolution is thus uncertain, it likely corresponds to a factor of 10 in luminosity, roughly the change seen in the ultraviolet luminosity density of the Universe as a whole, or a factor of order 300 in density. The evolution likely flattens off or possibly even reverses at higher redshifts.
6. Analysis of the relationship between the star-formation history of the Universe and the redshift distribution of the light in the background suggests that the observed redshift distribution in the sample is broadly consistent with a redshift evolution of the far-IR luminosity density that matches that in the ultraviolet, i.e. rising to a peak in the $1.2 < z < 2$ range and falling thereafter. This is supported by direct computation of the far-IR luminosity density. *The number of sources is still small, and the uncertainties large, but we can at least say that we see no strong evidence in the present data for significant differences in the redshift dependence.*
7. With the caveat that our estimates of the far-IR bolometric luminosities are based on the Arp 220 SED (and are thus uncertain by a factor of 2-3), the population of high redshift ULIRGs detected in the survey at $z > 1$ already appears to be producing a bolometric luminosity in the far-IR that matches that produced in the ultraviolet by the whole optically selected population. Thus these high redshift ULIRGs, and the lower luminosity systems presently below the detection threshold, are likely to be producing a significant fraction of all stars that have ever formed. Furthermore, they are doing so in systems with high star-formation rates, exceeding 300 $h_{50}^{-2}$ $M_O$ yr$^{-1}$, with indications of merger-like activity and in the presence of large amounts of dust (i.e. metals). *It is therefore attractive to identify these as forming the metal-rich spheroid population, in which case we would infer that much of this activity has occurred relatively recently, at $z \sim 2$.*

**Acknowledgements**

This program would not have been possible without the SCUBA instrument and we thank the many people who have designed and built this remarkable instrument. We applaud the national observatory system that has enabled us, along with a broad community, to have access to such innovative instruments. We thank the JCMT support staff, and especially Wayne Holland, for their assistance with this program. Gabriela Mallen-Ornelas kindly donated CFHT time for the measurement of the redshift of CFRS–10.1141. We benefited from conversations with John Peacock.. The scientific research of SJL and JRB is supported by the National Science and Engineering Research Council (NSERC) of Canada and by the Canadian Institute for Advanced Research (CIAR) and this support is gratefully acknowledged.

**Figure Captions**

*Fig 1:* *The distribution of P values derived for the optical identifications (from the optical plate material) compared with the distribution that would be expected if the optical and sub-mm populations were completely unrelated. The cut-off at P = 0.56 in both distributions is due to the finite search radius and density of optical galaxies at $I_{AB} < 25$, which produces a significant number of "empty fields" which are shown for presentation purposes at P > 1. The statistic P measures the chance that a particular sub-mm source would have an optical galaxy within the observed distance, but does not determine the chance that that optical identification is actually correct, since some low P values would be expected in a large enough sample. Thus the best measure of whether individual identifications are correct is given by comparing the number of identifications in the whole sample with a particular value of P with the number, NP, that would have been expected even if there was no physical association between the sub-mm and optical populations. The three identifications with lowest P are clearly very secure, the two with highest P are much less so – see text for details.*

.

*Fig 2:* *Montage of HST images of the 7 (of 8) securely identified sources in our sub-mm survey. These are arranged in order of ascending redshift. Each image shows an area 10×10 arcsec, except that for 03B (=CFRS–10.1411), a spiral galaxy at z = 0.074 which is 22×22 arcsec. The small insets for the four smallest galaxies show the central 1×1 $arcsec^2$ area magnified by a factor of three. All of the galaxies except CFRS–10.1411 show some evidence for disturbed morphologies and in some cases there is clear evidence for merger-like activity.*

*Fig 3:* *$(V-I)_{AB}/(I-K)_{AB}$ colors of 7 of 8 identifications compared with colors of template galaxies derived from Coleman et al (1980) (the light continuous lines show the tracks with redshift for E, Sbc and Irr, the dashed lines interpolations between them). The heavy solid lines*

*show the locus of galaxy types at z = 0.5, 1.5 and 2.5 respectively. A typical reddening vector for a simple $\lambda^{-1}$ reddening law is shown. The three sources with spectroscopically determined redshifts (as shown) have colours consistent with modest reddening..*

*Fig 4:* *Flux density at 850 µm plotted against measured or estimated redshift. Solid squares are three securely identified sources with spectroscopically determined redshifts, open squares are three securely identified sources with photometrically estimated redshifts, open circles are two probable identifications with estimated redshifts and the four stars are the four empty fields arbitrarily assigned redshifts z ~ 4 for display purposes. The lines show the flux density that Arp 220 (z = 0.018) would have as a function of redshift for $\Omega_0 = 1$ (solid) and $\Omega_0 = 0.2$ (dotted). All of the detected sources except the one at z < 0.1 have luminosities comparable or larger than that of Arp 220. There is no correlation between $S_{850}$ and redshift, as expected given the flat flux density-redshift relation at z > 0.5.*

*Fig 5:* *I-band magnitudes vs. redshift (measured or estimated) for the 8 identified sources and limits to the empty field sources (plotted at arbitrary redshift z ~ 4), compared with an unevolving Sbc galaxy from Coleman et al (1980) with present-day L\* luminosity. Symbols as in Fig 4. The identified galaxies are generally luminous L\* galaxies at visible wavelengths.*

*Fig 6:* *The (0.8-850) "color", defined as log ($S_{0.8}/S_{850}$) for the eight identifications and four empty fields, using the same symbols as in Figs 4 and 5. The points for the empty fields are upper limits. The lines show the colors of Arp 220 (solid line) and, in order of increasing $S_{0.8}/S_{850}$ ratio, the high extinction star-burst "SBH" , "LINER" and "spiral" spectral energy distributions from Schmitt et al (1997). The observed colors are consistent with those of Arp 220 at the estimated or measured redshifts of the galaxies. The empty field sources could*

*be similar sources at any redshifts z > 2, or more heavily extinguished sources at lower redshifts.*

*Fig 7:* *As in Fig 6 but for the (450-850) "color". Symbols with only one error bar are 2σ upper limits. Note how the four sources detected at 450 µm are 4 of the 5 sources with lowest measure or estimated redshifts, z < 1.5. All of the Schmitt (1997) SEDs are the same (long dashed line). Arp 220 is shown by solid line. Changing to a slightly lower emissivity, i.e. $\varepsilon \propto v^1$ would lower the Schmitt et al curves as indicated (dot-dash line).*

*Fig 8:* *As in Fig 6 and 7 but for the (15-850) "color" from ISO measurements and upper limits Flores et al 1998b). These are available in the CFRS-1417+52 field only and only one source is detected at both 15 µm and 850 µm. The remaining 850 µm sources are plotted as in Fig 5 and 6. The five crosses represent the lower limits to the colors of the sources detected at 15 µm but not at 850 µm, shown at their measured (open crosses) or estimated (simple crosses) redshifts. All of the sources are consistent with the colors of Arp 220 or the SBH SED of Schmitt et al (1997) at the measured or estimated redshifts.*

*Fig 9:* *As for Figs 6-8 except for the (850-5GHz) "color" - again only for sources in the CFRS-1417+52 field where radio data is currently available. As in Fig 8, crosses represent radio sources not detected at 850 µm (one is off the diagram at z = 0.010). All of the measurements or limits are broadly consistent with the Arp 220 SED, although the radio sources that are detected in the sub-mm are in fact also differentiated by radio spectral index from those that are not (see text).*

*Fig 10:* *(a) Above: The redshift distribution implied for our sample. Solid boxes are spectroscopically determined redshifts of secure identifications, hatched boxes are estimated redshifts of secure identifications and open boxes are estimated redshifts of "probable" identifications. The elevated region denotes the four "empty field" sources of*

*unknown redshift. Below: the redshift distribution for the HDF sample of Hughes et al (1998). Given the small numbers of objects, these appear to be consistent.*

*(b) Comparison of the distribution of I-band magnitudes for the sources in our sample (top) compared with those in the sample of gravitationally lensed sub-mm sources in Smail et al (1998) (middle) and the HDF sample of Hughes et al (1998). The "empty field" sources are shown as hatched boxes at their upper limits. These distributions are again broadly consistent.*

*Fig 11:*   *Cumulative bolometric luminosity function (CBLF). The point at $0.5 < z < 1$ is based on two objects, with Poisson error bars . The point at $1.0 < z < 3.0$ represents a minimum (4 sources) and a maximum (8 sources) depending on the redshifts of the empty field sources. Both minimum and maximum have associated Poisson errorbars. Finally, the point at $3 < z < 8$ is represented by a maximum (4 sources) with associated Poisson error-bar. The CBLF is compared with that derived from the IRAS 60 μm luminosity functions of Soifer et al (1987) and Saunders et al (1990), accounting for the different $H_0$ and definitions of $L_{bol}$ (see text), and with an ultraviolet CBLF at $z = 3$ from Dickinson (1998), for which, bolometric luminosities are computed simply as $\nu L_\nu$ at 1500 Å. The comparison of the bolometric luminosities in the two far-IR samples relies heavily on the assumption of an SED like that of Arp 220 and is correspondingly uncertain.. Nevertheless, substantial evolution is clearly required to $z \sim 1$ (see text for discussion) although it should be noted that this is less in luminosity than in number density. The evolution at higher redshifts is flatter and may indeed be negative.*

*Fig 12:*   *Estimated of the bolometric luminosity density in the ultraviolet (light symbols) and far-IR (heavy symbols). The treatment of the present sample mirrors in this Figure mirrors that in Fig 11. For simplicity, the error-bars are Poissonian based on the numbers of objects without taking into account their individual luminosities. The ultraviolet luminosity densities are "total estimates" integrated over most of the ultraviolet luminosity function, computing bolometric luminosity densities as $\nu \mathcal{L}_\nu$ using various estimates of $\mathcal{L}_\nu(z)$ over 1500 Å $< \lambda <$ 2800 Å (see text). The FIR bolometric luminosity density $\mathcal{L}_{FIR}$, is based on the Arp 220 SED and represents only luminosities above $2 \times 10^{12}$ $h_{50}^{-2}$ $L_0$ for $0.5 < z < 1.0$ and above $3 \times 10^{12}$ $h_{50}^{-2}$ $L_0$ at higher redshifts. Addition of lower luminosity sources will likely raise these points by a factor of up to 5. Readers should refer to the text for important caveats regarding the interpretation of this diagram.*

*Fig 13:*   *Production of the 850 μm background from different star formation histories: The left hand panel shows different heuristic star-formation histories, ranging from "Madau-type" curves peaking at $z = 1.2$ and 2, a model that rises to $z = 2$ and then has constant star-formation, a model that produces half the stars prior to $z = 3$ and finally a model with constant SFR at all epochs. For each model, the cumulative production of stars is shown in the center panel, and the cumulative distribution of light in the 850 μm background is shown in the right-hand panel. Because of the highly beneficial k-corrections at 850 μm, the light in the background is heavily weighted in favour of high redshift star-formation (see text for details). Models in which half the obscured star-formation in the Universe occurred prior to $z \sim 3$ predict that only 15% of the background should come from $z < 3$. We can rule this out unless all of the sources below the sample limit with $S_{850} < 3$ mJy have $z > 3$ – an unlikely situation. The irregular line in the left-most panel shows the distribution in redshift of the background produced by the observed sources – illustrating the effect of assuming that the sources below the limit of the survey have the same redshift distribution.*

**Table 1: Source identifications**

| Submm source[a] | CFRS[b] | FKWW[c] | RA[2000][d] | δ[2000][d] | $r^e$ 850/450 | P | P'[d] |
|---|---|---|---|---|---|---|---|
| 10B | 10.1411 | | 10 00 37.21 | +24 14 59.7 | 2.8/1.8 | $<10^{-4}$ | $<10^{-3}$ |
| 14F | 14.1139 | | 14 17 42.04 | +52 30 25.7 | 2.0/0.6 | $<10^{-3}$ | 0.002 |
| | | 15V.24 | 14 17 42.09 | +52 30 25.2 | 2.1/1.3 | $<10^{-3}$ | $<10^{-3}$ |
| 14A | … | | 14 17 40.21 | +52 29 06.5 | 1.7/…. | 0.05 | 0.18 |
| | | 15V.18 | 14 17 40.33 | +52 29 05.9 | 2.1/…. | $10^{-3}$ | $10^{-3}$ |
| 10A | 10.1153 | | 10 00 38.27 | +25 14 50.4 | 1.2/0.8 | $10^{-3}$ | 0.007 |
| 03A | 03.1054 | | 03 02 36.35 | +00 08 11.4 | 2.8/2.0 | 0.04 | 0.14 |
| 10D | 10.1167 | | 10 00 36.86 | +25 14 44.0 | 2.8/…. | 0.04 | 0.15 |
| 14C | … | | 14 17 33.81 | +52 30 49.9 | 2.1/…. | 0.07 | 0.22 |
| 14B | … | | 14 17 51.92 | +52 30 27.5 | 1.9/…. | 0.14 | 0.32 |

Notes to Table 1:
[a] As catalogued in Paper 1
[b] CFRS catalogue number if catalogued
[c] FKWW radio source (from Fomalont et al 1992)
[d] Position of optical or radio identification
[d] Positional offset from nominal sub-mm position (second figure for 450 μm where available)
[e] Corrected probability that identification is not associated with sub-mm source.

**Table 2: Photometry of identifications**

| Submm source[a] | CFRS[b] | $z^c$ | $U_{AB}$ | V | $I_{AB}$ | $K_{AB}$ | $S_{5GHz}$[e] μJy | $S_{7\mu m}$[f] μJy | $S_{15\mu m}$[g] μJy |
|---|---|---|---|---|---|---|---|---|---|
| 10B | 10.1411 | 0.074 | … | 18.2 ± 0.1 | 17.3 ± 0.1 | 16.4 ± 0.1 | … | … | … |
| 14F | 14.1139 | 0.660 | 22.7 ± 0.1 | 21.6 ± 0.1 | 20.3 ± 0.1 | 18.6 ± 0.1 | 79 ± 6 | 115 ± 34 | 562 ± 50 |
| 14A | … | (2.0) | 26.5 ± 0.5 | 25.5 ± 0.8 | 24.1 ± 0.3 | 20.8 ± 0.1 | 44 ± 4 | < 150 | < 200 |
| 10A | 10.1153 | 0.550 | … | 23.3 ± 0.1 | 21.2 ± 0.1 | 18.9 ± 0.1 | … | … | … |
| 03A | 03.1054 | (1.5) | 25.8 ± 0.4 | 24.5 ± 0.4 | 23.4 ± 0.2 | … | … | … | … |
| 10D | 10.1167 | (0.8) | … | 24.1 ± 0.3 | 22.6 ± 0.1 | 20.5 ± 0.2 | … | … | … |
| 14C | … | (2.5) | 25.7 ± 0.3 | 24.8 ± 0.3 | 23.8 ± 0.2 | 21.1 ± 0.1 | < 16 | < 150 | < 200 |
| 14B | … | (2.5) | 27.6 ± 0.8 | 26.0 ± 0.5 | 25.2 ± 0.4 | >22 (2σ) | < 16 | < 150 | < 200 |

Notes to Table 2:
[a] As catalogued in Paper I and Table I
[b] CFRS catalogue number if catalogued
[c] Redshift. Bracketed are crude photometric estimated from UVIK photometry..
[e] from Fomalont et al (1992) and Hammer et al (1995b).
[f] from Flores et al (1998a)
[g] from Flores et al (1998b)

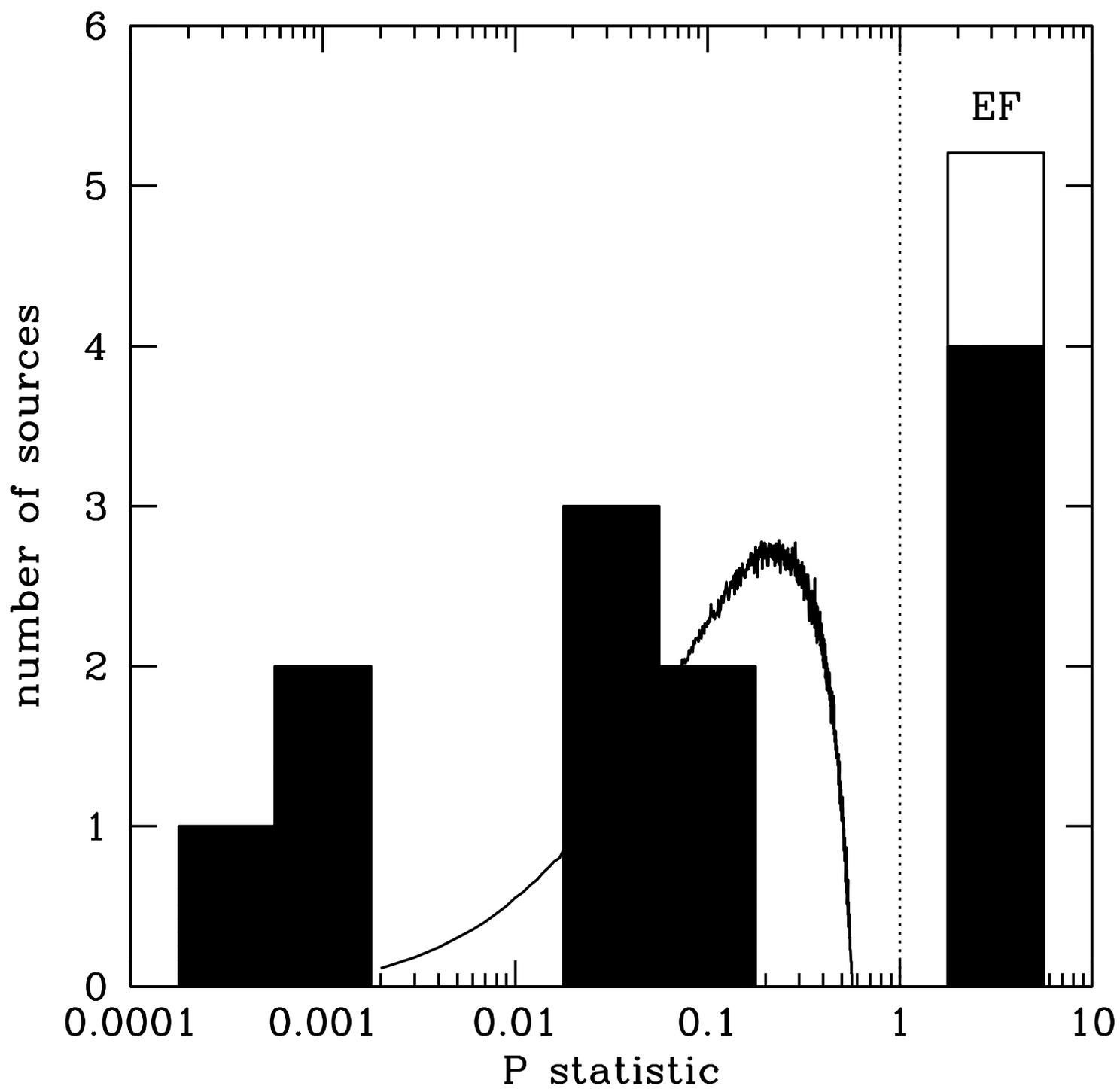

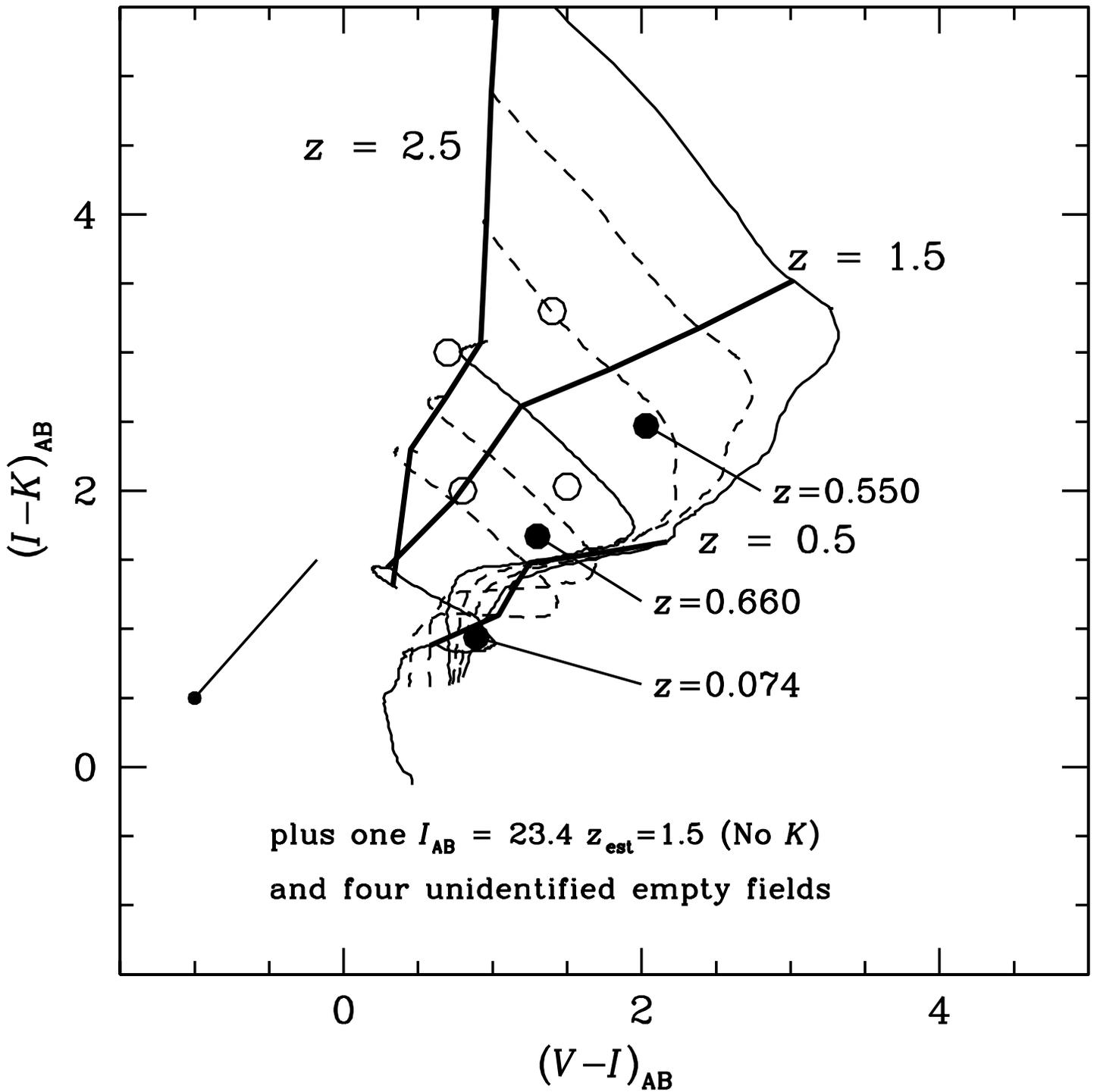

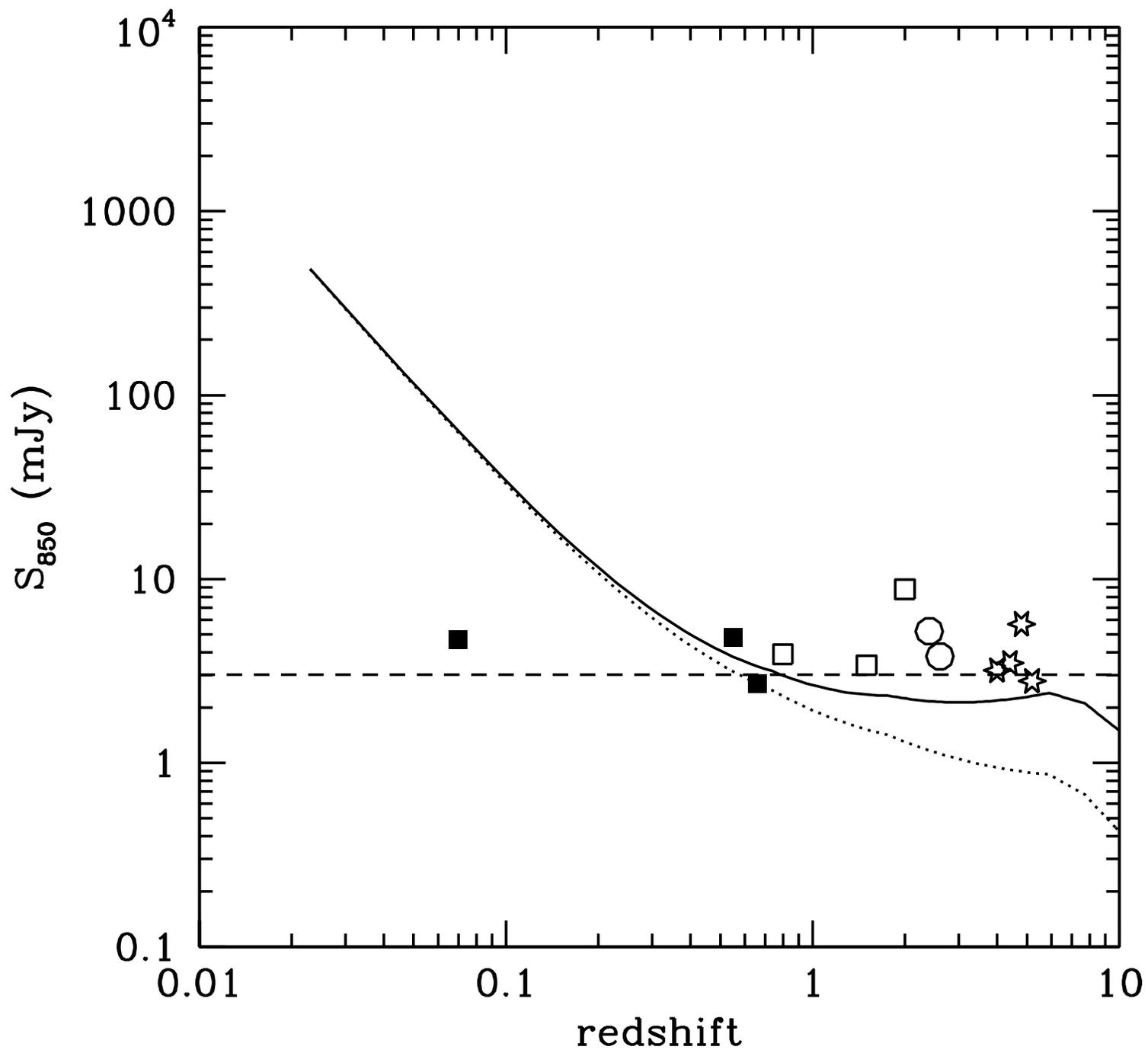

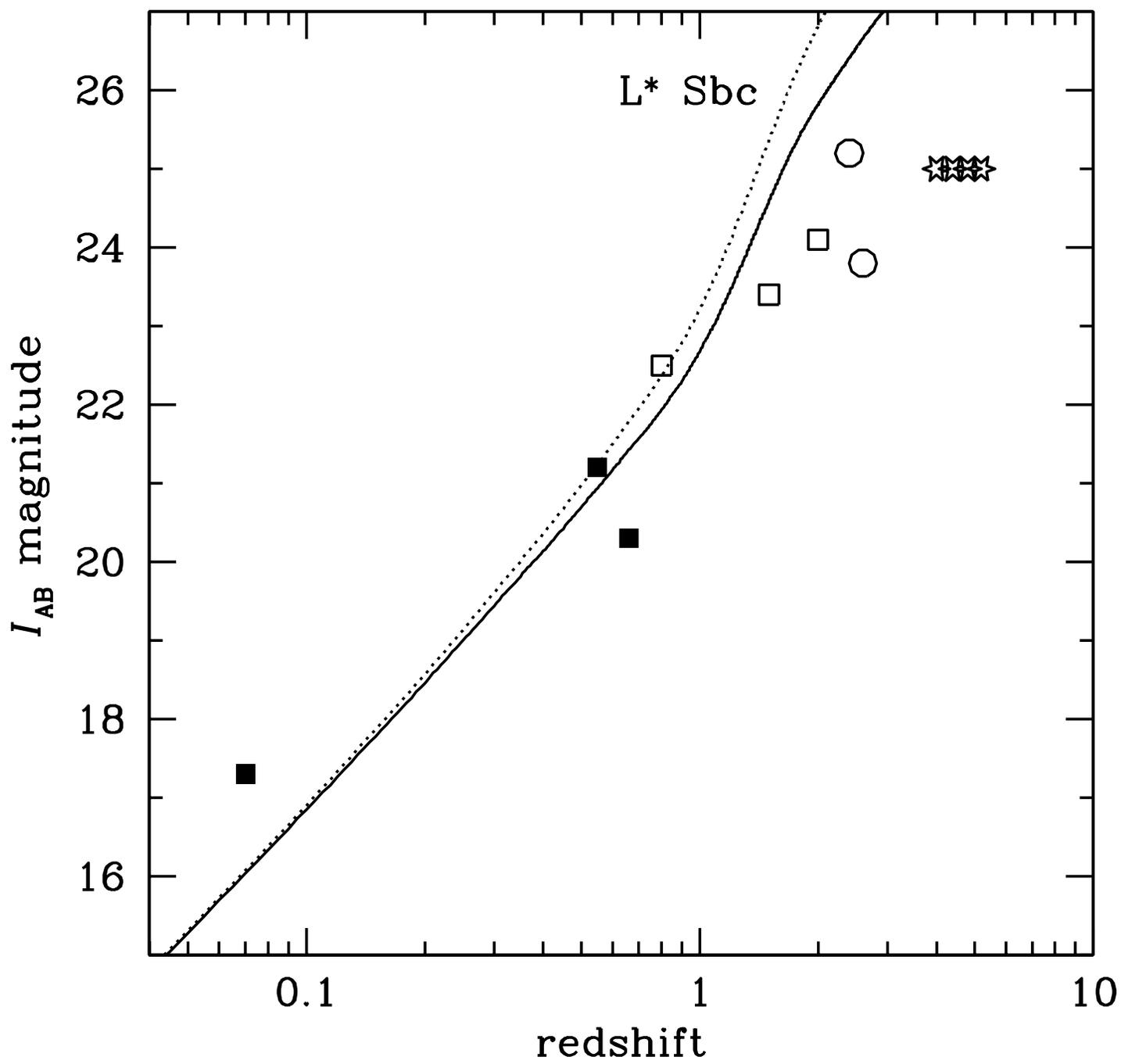

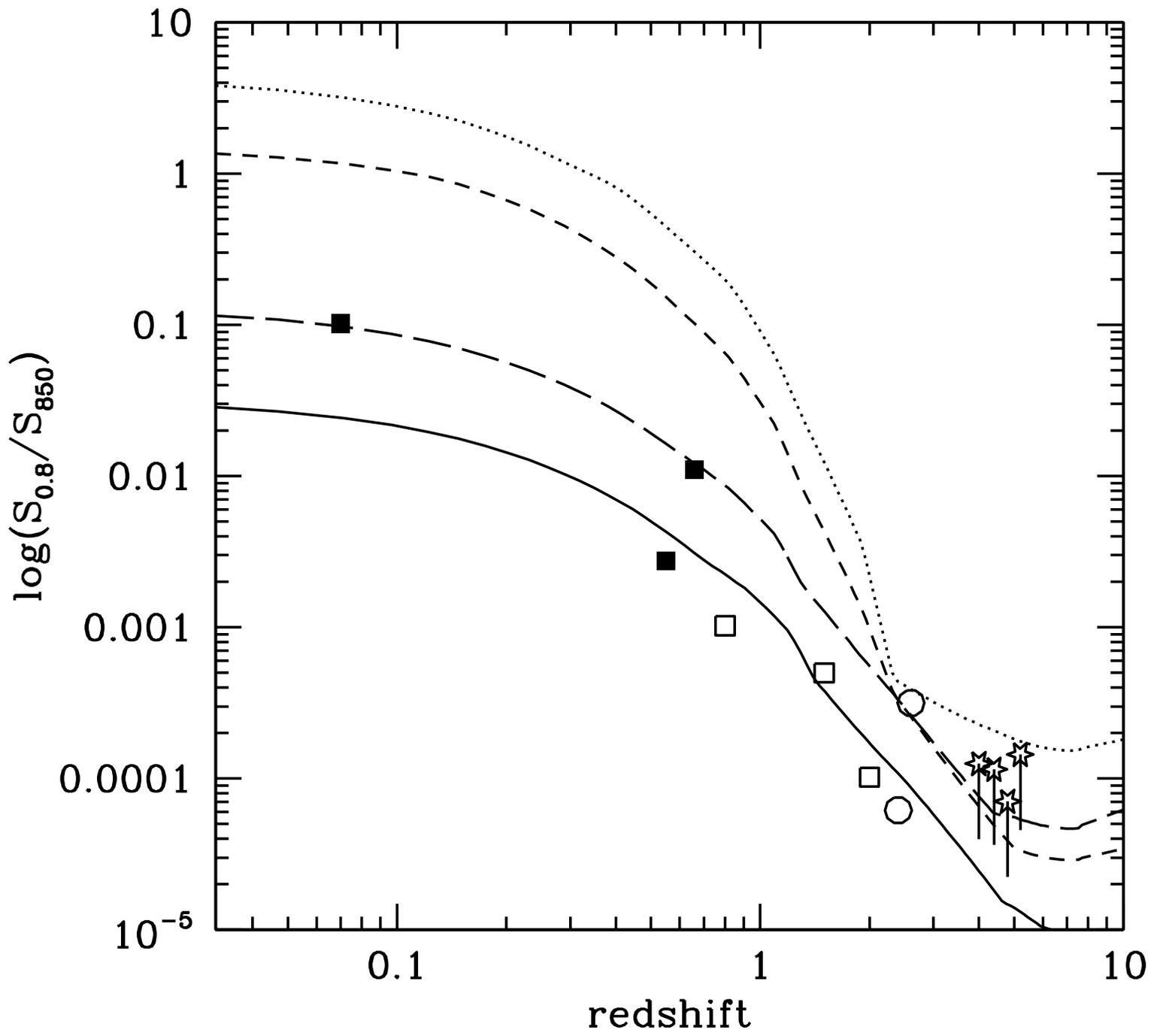

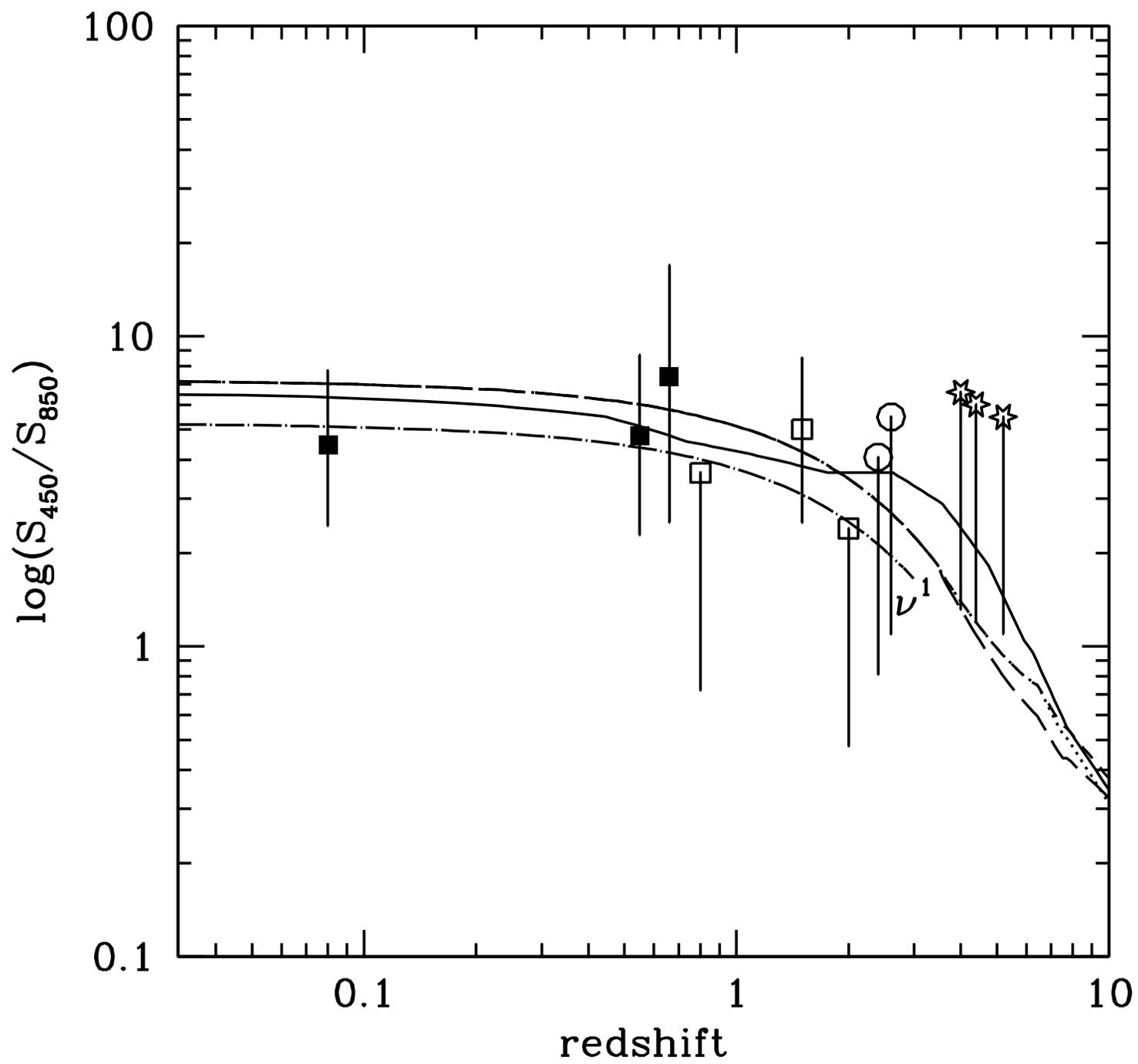

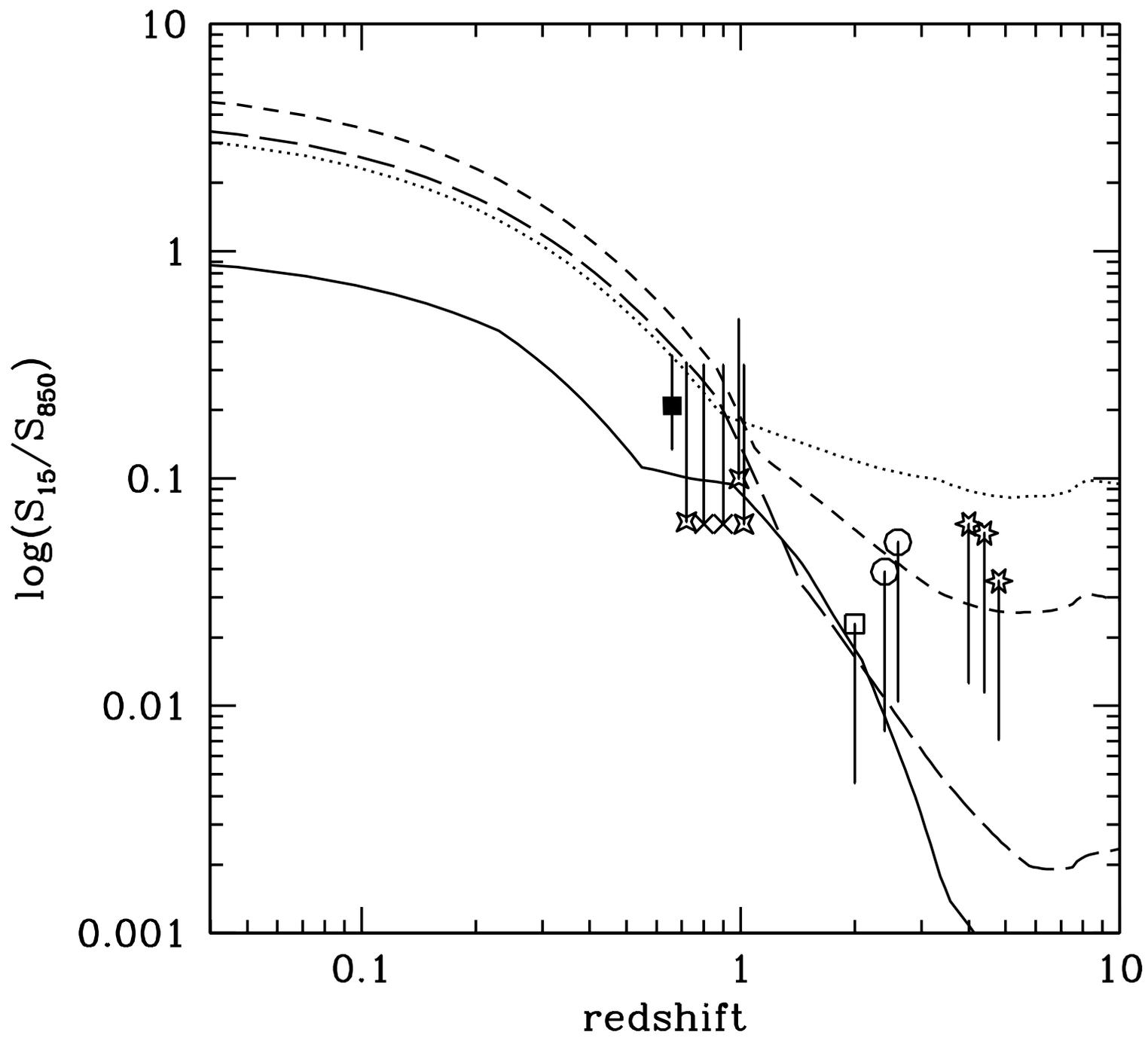

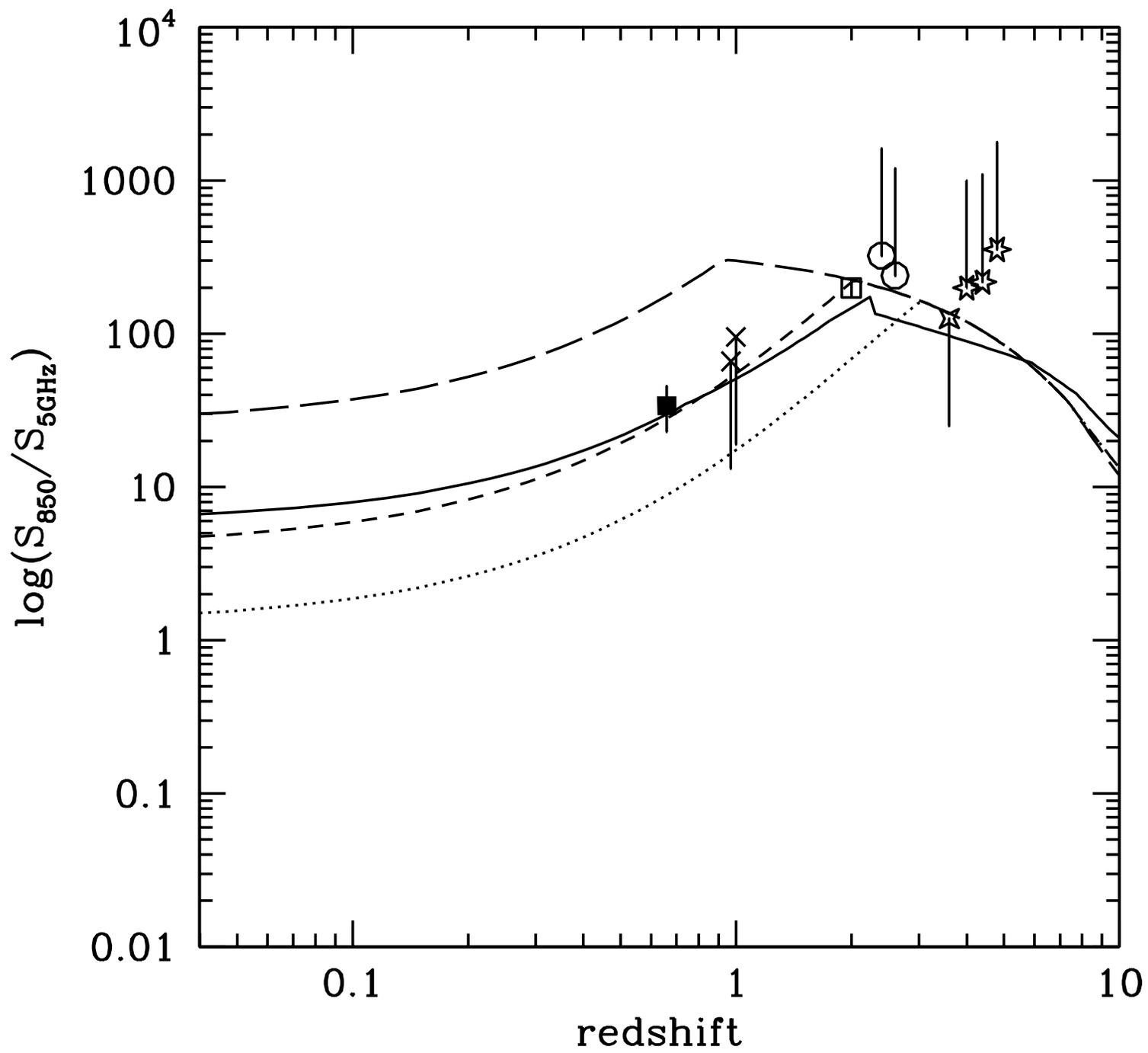

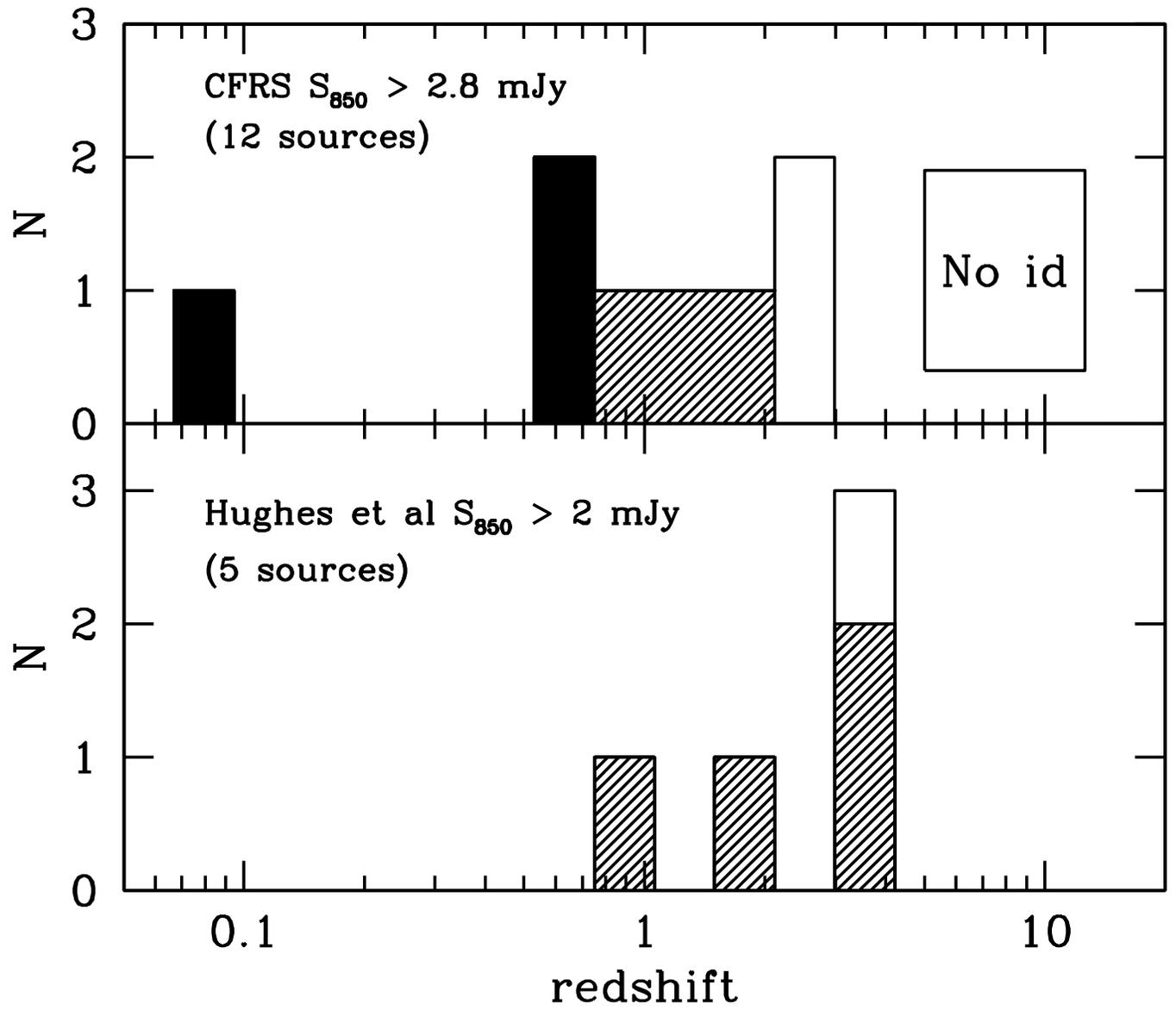

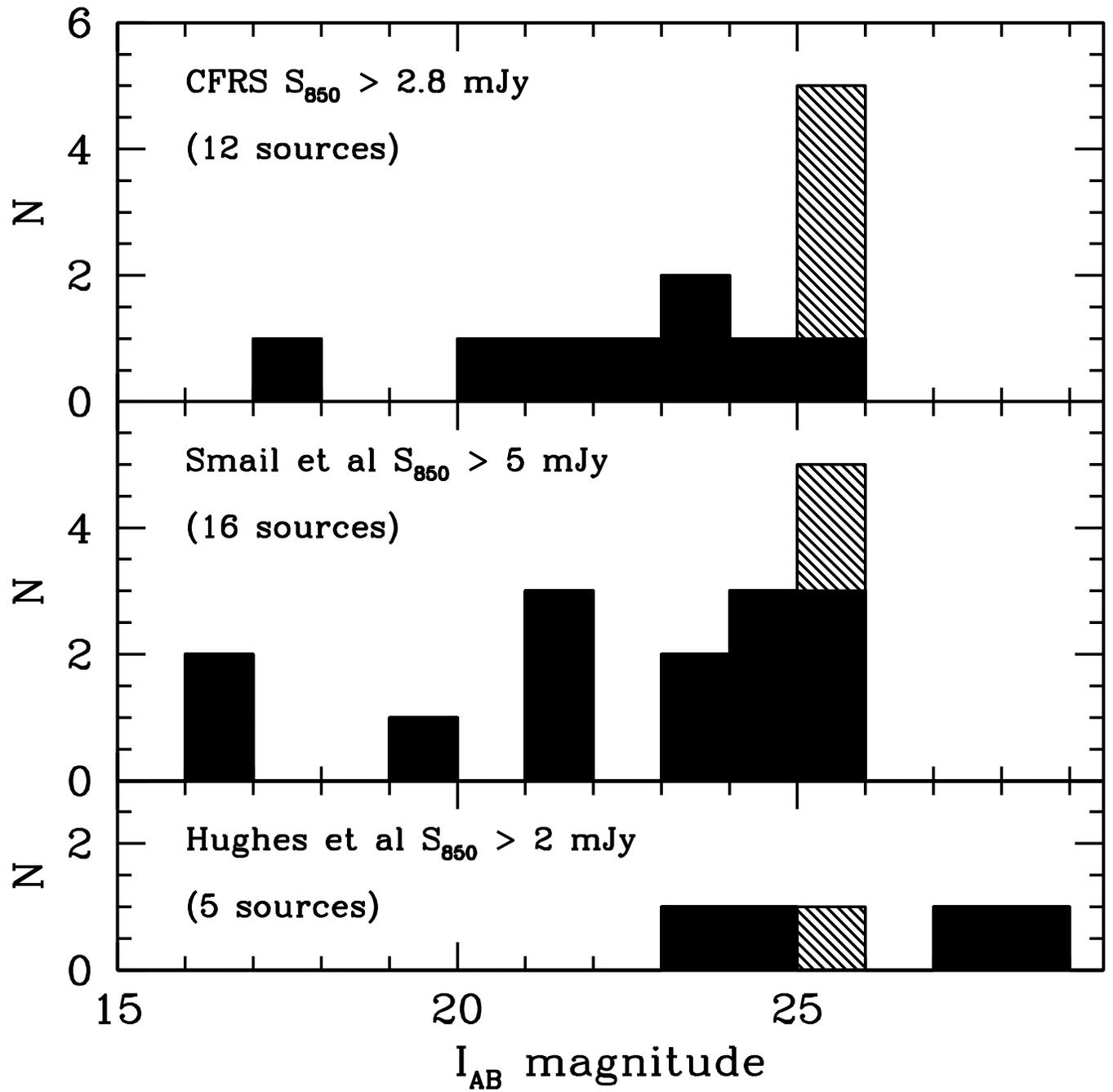

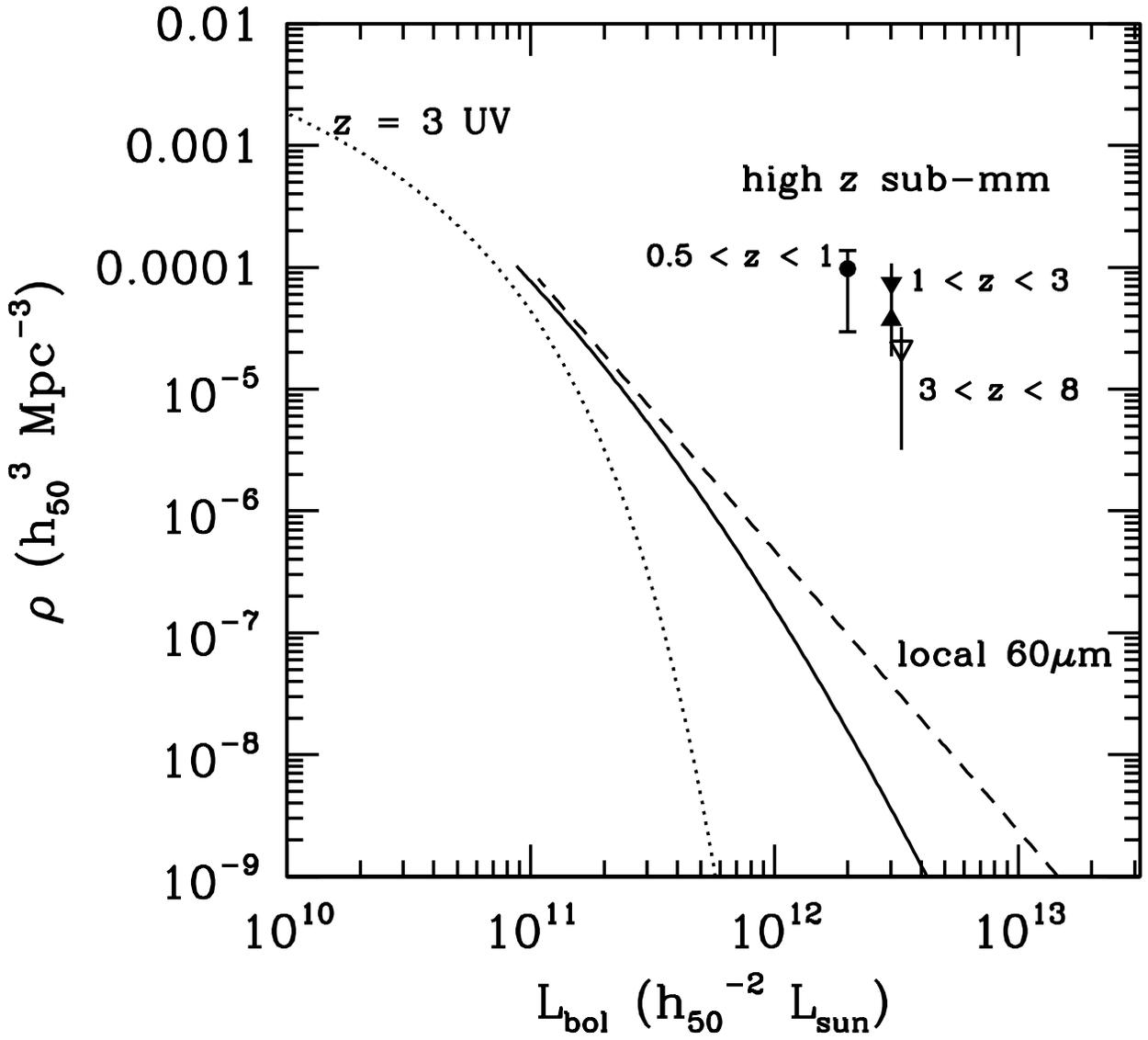

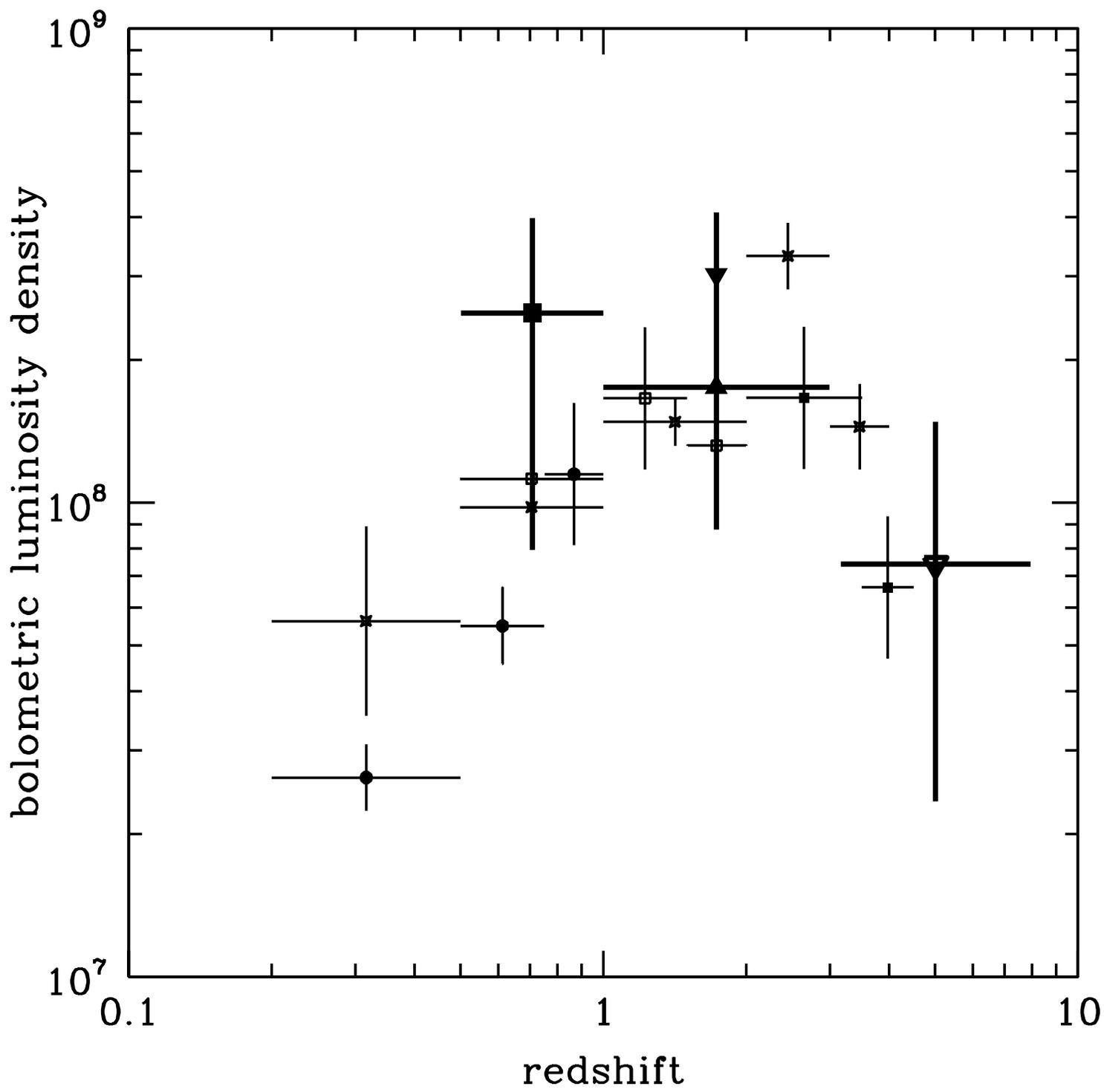

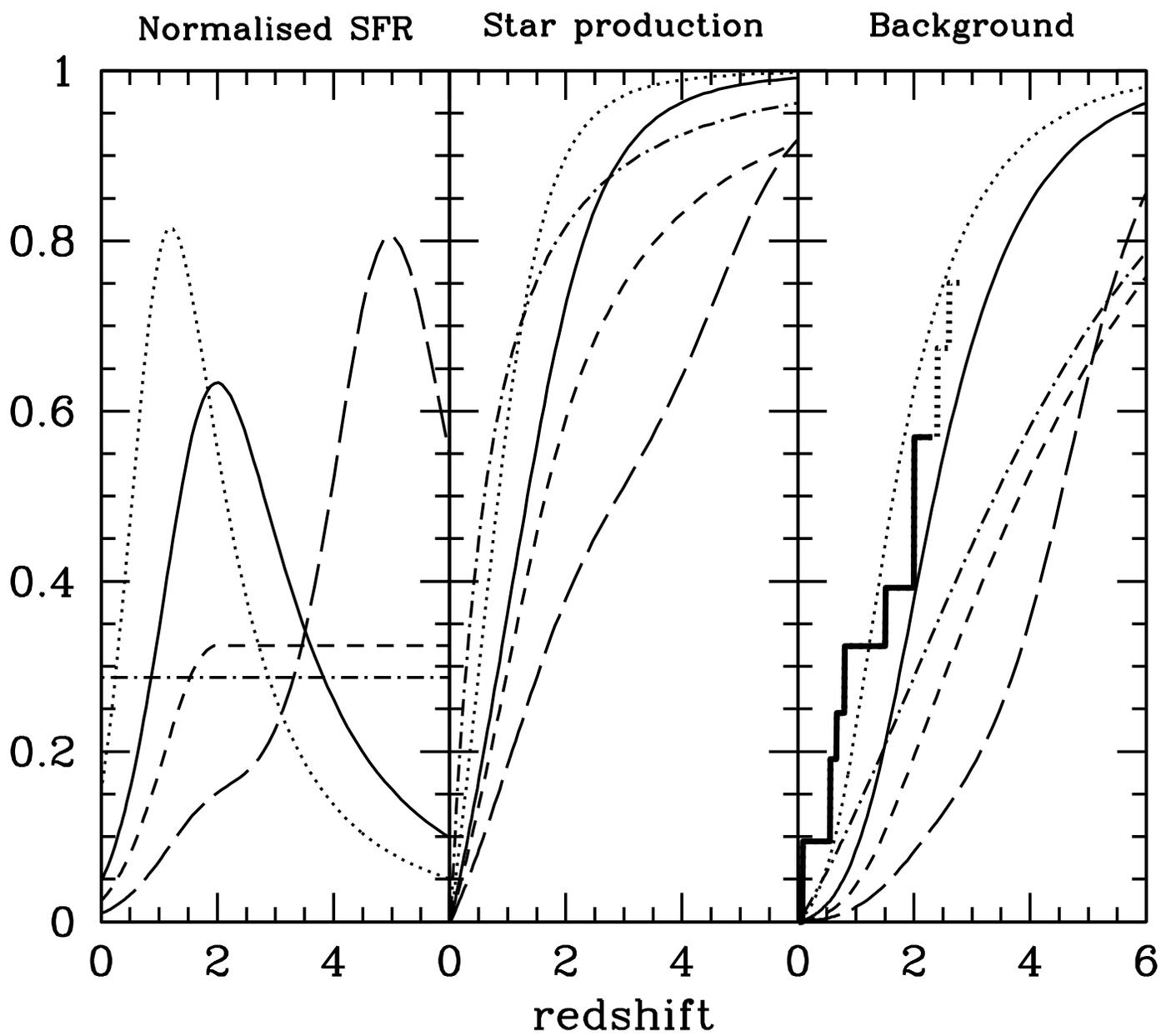